\pdfoutput=1

\documentclass[11pt,twoside,a4paper,cmspaper,final,collab]{cms-tdr}
\def\svnVersion{231468}\def\cmsCernNo{2013-224}\def\cmsCernDate{\today}\def\cmsMessage{Published in the Journal of High Energy Physics as \href{http://dx.doi.org/10.1007/JHEP03(2014)032}{\doi{10.1007/JHEP03(2014)032.}}}

\begin{document}\cmsNoteHeader{FSQ-12-028}

\hyphenation{had-ron-i-za-tion}
\hyphenation{cal-or-i-me-ter}
\hyphenation{de-vices}

\RCS$Revision: 226202 $
\RCS$HeadURL: svn+ssh://svn.cern.ch/reps/tdr2/papers/FSQ-12-028/trunk/FSQ-12-028.tex $
\RCS$Id: FSQ-12-028.tex 226202 2014-02-05 21:42:33Z alverson $
\providecommand{\rj}{\ensuremath{\mathrm{j}}\xspace}
\providecommand{\MT}{\ensuremath{M_{\mathrm{T}}}\xspace}
\newcommand{\DS}{\ensuremath{\Delta\mathrm{S}}\xspace}
\cmsNoteHeader{FSQ-12-028} 
\title{Study of double parton scattering using W + 2-jet events in proton-proton collisions at $\sqrt{s} = 7$\TeV}

\date{\today}

\abstract{
Double parton scattering is investigated in proton-proton collisions at $\sqrt{s} = 7$\TeV where the final state includes a W boson, which decays into a muon and a neutrino, and two jets. The data sample corresponds to an integrated luminosity of 5\fbinv, collected with the CMS detector at the LHC. Observables sensitive to double parton scattering are investigated after being corrected for detector effects and selection efficiencies.
The fraction of W + 2-jet events due to double parton scattering is measured to be  $0.055 \pm 0.002\stat \pm 0.014\syst$. The effective cross section, $\sigma_\text{eff}$, characterizing the effective transverse area of hard partonic interactions in collisions between protons is measured to be $20.7 \pm 0.8\stat\pm  6.6\syst\unit{mb}$.
}

\hypersetup{%
pdfauthor={CMS Collaboration},%
pdftitle={Study of double parton scattering using W + 2-jet events in proton-proton collisions at sqrt(s) = 7 TeV},%
pdfsubject={CMS},%
pdfkeywords={CMS, physics, QCD, jet physics}}

\maketitle 

\section{Introduction}

In high-energy proton-proton (pp) collisions at the Large Hadron Collider (LHC), semi-hard parton-parton scattering, producing particles with transverse momenta \PT of a few \GeVns, dominates the inelastic cross section.
In such processes longitudinal momentum fractions, given by $x \sim 2 \PT/\sqrt{s}$, of values down to $\mathcal{O}(10^{-3})$ are probed.
 At these values of $x$, the parton densities are large causing a sizable probability for two or more parton-parton scatterings within the same pp interaction~\cite{Sjostrand:1986ep}.
 Such multi-parton interactions (MPI) at semi-hard scales of a few \GeVns{}s have been observed in high-energy hadronic collisions~\cite{Bartalini:2010su}.
 Conversely, the evidence for hard double parton scattering (DPS) processes in the same pp collision at scales of a few tens of \GeVns is still relatively weak.
 In processes where a W and two jets are produced, the $x$ values are larger, $x \sim 10^{-2}$, and the parton densities are lower.
 However, a sizable contribution to DPS can still be expected if the second scattering, yielding two jets, occurs at a high rate.
 The study of DPS processes is important because it provides valuable information on the transverse distribution of partons in the proton~\cite{Diehl:2011yj} and on the multi-parton correlations in the hadronic wave function~\cite{Calucci:2010wg}.
 DPS also constitutes a background to new physics searches at the LHC~\cite{DelFabbro:1999tf,Hussein:2006xr,Bandurin:2010gn}.

\begin{figure}[htbp]
\centering
\includegraphics[width=0.40\textwidth]{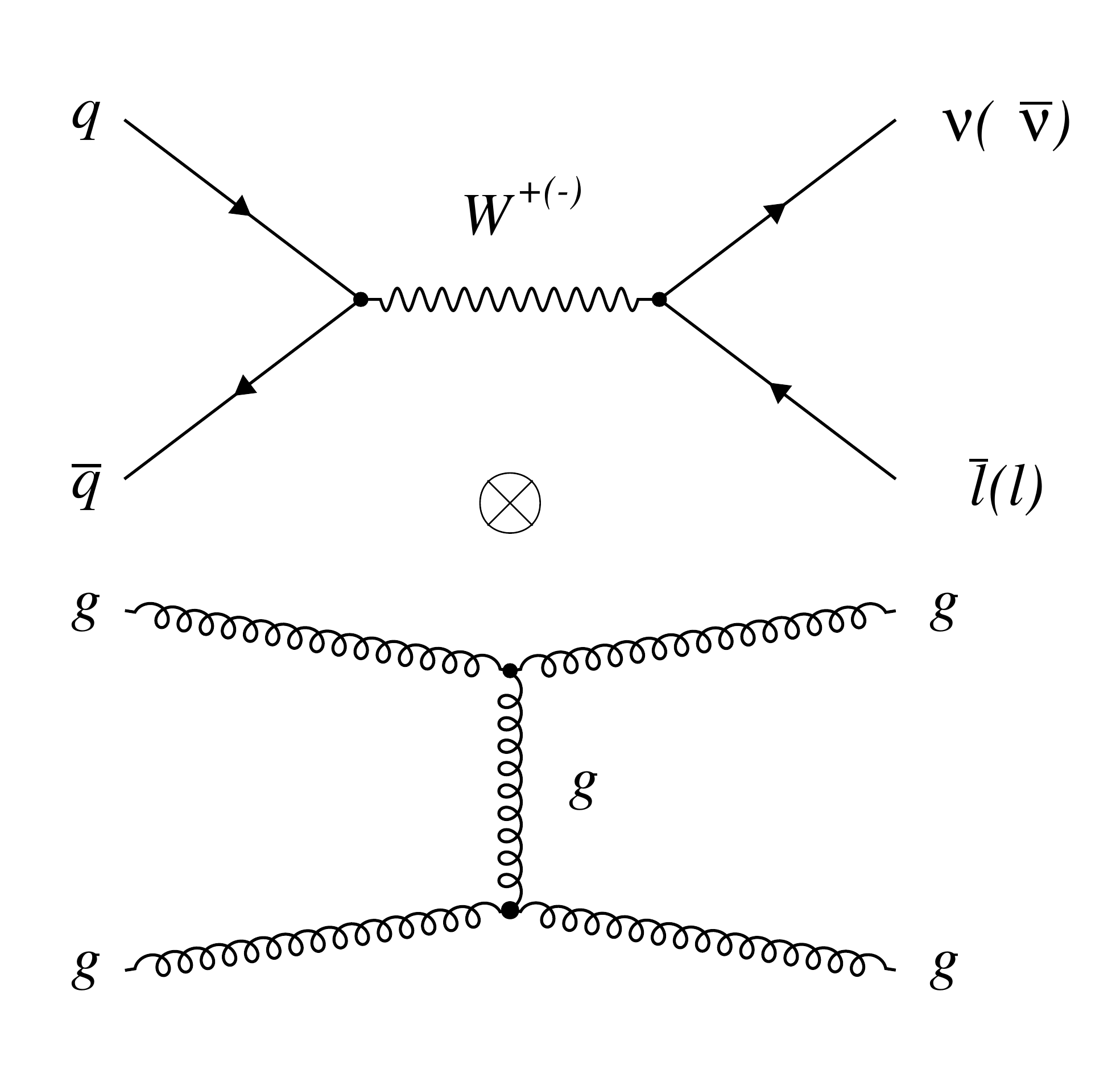}
 \includegraphics[width=0.40\textwidth]{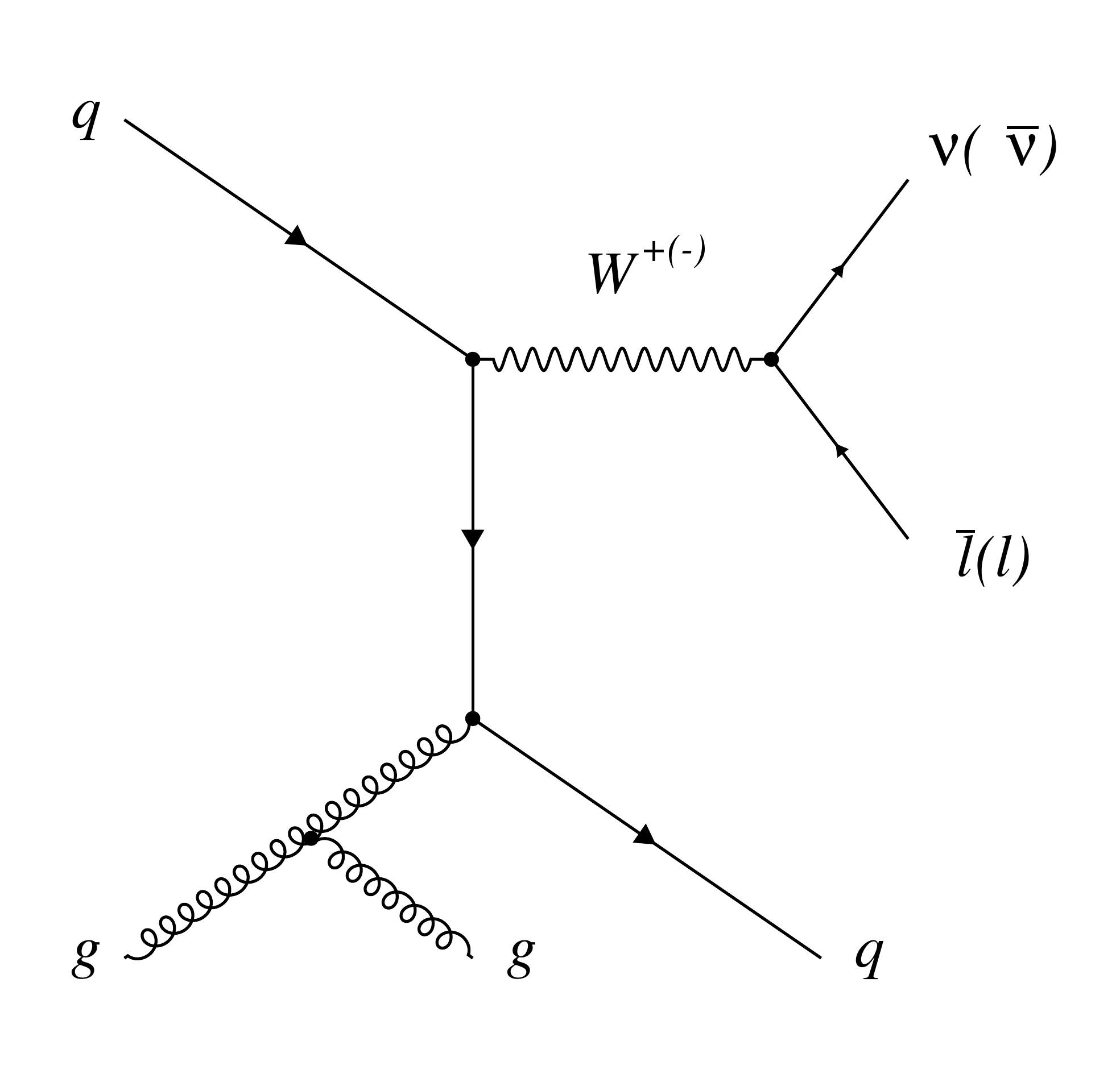}
 \caption{Feynman diagrams for W + 2-jet production from (left) double parton scattering and (right) single parton scattering.}
\label{fig:feyn}
\end{figure}

Various measurements in pp and $\Pp\Pap$ collisions at $\sqrt{s}=63\GeV$~\cite{UA2}, 630\GeV~\cite{AFS}, and 1.8\TeV~\cite{CDF1} are consistent with DPS contributions to multijet final states, as well as to $\gamma$ + 3-jet events at $\sqrt{s}=1.8\TeV$~\cite{CDF2} and 1.96\TeV~\cite{D0}.
Additional searches for DPS have been proposed via double Drell--Yan, four jet, and same-sign WW production, as well as in W production associated with jets~\cite{dyDPS1,dyDPS2,jetDPS1,jetDPS2,Berger:2009cm,wjetDPS1,Berger:2011ep,wwDPS1}.
This paper presents a study of DPS based on W + 2-jet events in pp collisions at 7\TeV.
DPS with a W + 2-jet final state occurs when one hard interaction produces a W boson and another produces a dijet in the same pp collision, as sketched in Fig.~\ref{fig:feyn}(left).
The W + 2-jet process is attractive because the muonic decay of the W provides a clean tag and the large dijet production cross section increases the probability of observing DPS.
Events containing a W + 2-jet final state originating from single parton scattering (SPS) constitute an irreducible background (Fig.~\ref{fig:feyn}(right)).
The ATLAS Collaboration has carried out a similar DPS measurement using W + 2-jet events at $\sqrt{s}=7\TeV$~\cite{atlas}.

The outline of this paper is as follows. Section 2 describes DPS in terms of effective cross section and defines the relevant observables.
Section 3 presents a brief description of the Compact Muon Solenoid (CMS) detector, the data and the simulated samples, as well as the event selection criteria.
Section 4 summarizes the unfolding of the DPS-sensitive observables, the systematic effects, and the comparison of data and simulation.
The method to extract the DPS fraction is discussed in Section 5.
Section 6 presents the extraction of the DPS fraction from the data and corresponding systematic uncertainties.
The measurement of the effective cross section is described in Section 7.

\section{Effective cross section}
\label{sec:effxsec}
The effective cross section, $\sigma_\text{eff}$, is a measure of the transverse distribution of partons inside the colliding hadrons and their overlap in a collision.
 The effective cross section involves the cross section for two processes to occur simultaneously and the cross sections for the individual processes.
 If A and B are two independent processes, whose production cross sections are $\sigma_\mathrm{A}$ and $\sigma_\mathrm{B}$, respectively, $\sigma_\text{eff}$ can be written as:

\begin{equation}
\sigma_\text{eff} = \frac{m}{2}\frac{\sigma_\mathrm{A}\cdot\sigma_\mathrm{B}}{\sigma^{\mathrm{DPS}}_\mathrm{A+B}},
\end{equation}

where ``$m$'' is a symmetry factor for indistinguishable ($m = 1$) and distinguishable ($m = 2$) final-states
and $\sigma^{\mathrm{DPS}}_\mathrm{A+B}$ is the cross section of the two processes to occur simultaneously.

According to various phenomenological studies~\cite{Treleani, Seymour1, Seymour2}, the above cross sections should be inclusive.
This requirement makes the determination of $\sigma_\text{eff}$ independent of the specific mechanisms of the first and second interactions, as well as of the parton distribution functions (PDF).
Inclusive $\sigma^{\mathrm{DPS}}_\mathrm{A + B}$ also includes contributions from higher number of parton scatters.

However, in the present analysis an exclusive selection is performed by considering the events with one W boson and exactly two jets with $\pt> 20$\GeVc and pseudorapidity, $\eta$, within $\pm$2.
The pseudorapidity is defined as $\eta = -\ln [ \tan(\theta/2)]$, where $\theta$ is the polar angle measured with respect to the anti-clockwise beam direction.
This sample should have a significant contribution from events where one interaction produces only a W boson and no jet with $\pt> 20$\GeVc within $\abs{\eta} < 2.0$ and the other interaction produces exactly two jets with $ \pt> 20$\GeVc and $\abs{\eta} < 2.0$.

Experimentally, the exclusive selection is necessary to identify the jets from the second interaction.
In this case, DPS-sensitive observables can be defined based on the back-to-back topology of the two jets.
From a sample of simulated events generated with \MADGRAPH5~\cite{Alwall:2011,Maltoni:2003} followed by hadronization and parton showering (PS) using the 4C tune~\cite{Corke:2010yf} of \PYTHIA8~\cite{Sjostrand:2007gs}, it is found that $\sigma_\text{eff}$ changes by only 2--3\% if an inclusive selection is applied.

In order to account for missing contributions of a larger number of parton scatterings, corrections~\cite{Treleani, Seymour1, Seymour2} were proposed to a previous DPS measurement of CDF~\cite{CDF2}.
However, in the kinematic region of the present study, due to the requirement of having exactly 2 jets, the contribution of triple and higher number of scatters is expected to be small and is estimated, with the same sample of simulated events as mentioned above, to be less than 1\% of the DPS contribution. 
Therefore, for the present analysis no additional correction is required for the exclusive selection.

Assuming independent interactions from DPS, $\sigma_\text{eff}$ can be rewritten in terms of the cross sections at the stable particle level (defined as lifetime, $c\tau > 10$ mm) within the detector acceptance.
For the case of the W + 2-jet process, $\sigma_\text{eff}$ becomes:

\begin{equation}
 \sigma_\text{eff} = \frac{\sigma'_{\PW + 0\rj}}{\sigma^{\prime{\mathrm{DPS}}}_{\PW+2\rj}}\cdot \sigma'_{2\rj} ,
\end{equation}

where the prime indicates that the cross sections are obtained at particle level.
The $\sigma'_{\PW + 0\rj}$ and $\sigma'_{2\rj}$ are the particle-level cross sections for W-boson production associated with zero-jet and  for dijet events, respectively.
The particle-level cross section for DPS events producing a W + 0-jet from the first interaction and exactly two jets from the second interaction is denoted by $\sigma'^{\mathrm{DPS}}_{\PW + 2\rj}$.
The cross sections $\sigma'_{\PW + 0\rj}$ and $\sigma'^{\mathrm{DPS}}_{\PW + 2\rj}$ are extracted from the same data sample; therefore $\sigma_\text{eff}$ can be reformulated in terms of the yield of W bosons associated with zero jets and the yield associated with DPS:

\begin{equation}
 \sigma_\text{eff} = \frac{N'_{\PW + 0\rj}}{N'^{\mathrm{DPS}}_{\PW + 2\rj}}\cdot \sigma'_{2\rj} .
\end{equation}

If we define the DPS fraction as

\begin{equation}
{f_\mathrm{DPS}} = \frac{N'^{\mathrm{DPS}}_{\PW + 2\rj}}{N'_{\PW + 2\rj}} ,
\end{equation}

 $\sigma_\text{eff}$ can be written as

\begin{equation}
 \sigma_\text{eff} = \frac{N'_{\PW + 0\rj}}{ f_\mathrm{DPS} \cdot N'_{\PW + 2\rj}}\cdot \sigma'_{2\rj},
\end{equation}
 or
 \begin{equation}
 \sigma_\text{eff} = \frac{R}{f_\mathrm{DPS}}\cdot \sigma'_{2\rj} ,
\end{equation}

where $R = N'_{\PW + 0\rj}/N'_{\PW + 2\rj}$. Thus, the determination of the effective cross section reduces to a measurement of R,  $\sigma'_{2\rj}$, and ${f_\mathrm{DPS}}$.

For the extraction of the DPS fraction, ${f_\mathrm{DPS}}$, observables that can discriminate between SPS and DPS are needed.
For DPS events, the W and the dijet system are independent of each other, while for SPS events they are highly correlated.
It is thus possible to define several observables that discriminate between DPS and SPS events.
The present analysis uses the following observables, which were also considered in previous DPS measurements at the LHC and
the Tevatron:

\begin{itemize}
\item
the relative ${\pt}$-balance between the two jets, $\Delta^\text{rel}~{\pt}$, defined as:

\begin{equation}
\Delta^\text{rel}~{\pt} = \frac{ \abs{{\ptvec({\rj_{1}})} + {\ptvec({\rj_{2}})} } }{\abs{{\ptvec({\rj_{1}})}} + \abs{{\ptvec({\rj_{2}})} } }.
\end{equation}

Here ${\ptvec({\rj_{1}}) }$ and ${\ptvec({\rj_{2}})}$ are the transverse momentum vectors of the leading (in ${\pt}$) and subleading jets.
In DPS events, at leading order (LO), the two jets balance each other and  $\Delta^\text{rel}~\pt$ is small, which is not the case for SPS events.
\item
The azimuthal angle between the W-boson and the dijet system, $\DS$, defined as:

\begin{equation}
\DS = \arccos \left( \frac{ \ptvec(\mu, \ETslash)\cdot\ptvec(\rj_1, \rj_2)}{\abs{\ptvec(\mu, \ETslash)}\cdot\abs{\ptvec(\rj_1, \rj_2)}} \right),
\end{equation}
where ${\ptvec(\mu, \ETslash)}$ and ${\ptvec({\rj_1}, {\rj_2})} $ are the combined transverse momentum vectors of ($\mu, \ETslash$) and the two jets, respectively, with $\ETslash$ as the missing transverse energy in the event, which is a measure of the transverse energy carried away by the neutrino from the W-boson decay.  In DPS events, the W and dijet momentum vectors are randomly oriented, whereas in SPS events the W and the dijet momenta vectors tend to be back-to-back at LO.

\end{itemize}

\section{Experimental methods}
\label{sec:expmethod}
In the present analysis a sample of W + 2-jet events is selected from a data sample of pp collisions at $\sqrt{s} = 7$\TeV with the CMS detector. 
The data sample corresponds to an integrated luminosity of 5\fbinv.
The distributions are fully corrected for detector effects and efficiencies.
These distributions are used for the extraction of the DPS fraction and the determination of the effective cross section.
The dijet production cross section required for the determination of the effective cross section is measured with a pp data sample collected in 2010 also at $\sqrt{s} = 7$\TeV. This sample corresponds to an integrated luminosity of 35\pbinv.

The central feature of the CMS apparatus is a superconducting solenoid of 6\unit{m} internal diameter, providing a magnetic field of 3.8\unit{T}.
 Within the superconducting solenoid volume are a silicon pixel and strip tracker, a lead tungstate crystal electromagnetic calorimeter (ECAL), and a brass/scintillator hadron calorimeter (HCAL).
 Muons are measured in gas-ionization detectors embedded in the steel flux return yoke outside the solenoid.
In addition, CMS has extensive forward calorimetry.
The CMS experiment uses a right-handed coordinate system, with the origin at the nominal interaction point, the $x$ axis pointing to the centre of the LHC ring, the $y$ axis pointing up (perpendicular to the plane of the LHC ring), and the $z$ axis along the anticlockwise-beam direction. The polar angle $\theta$ is measured from the positive $z$ axis and the azimuthal angle $\phi$ is measured in the $x$-$y$ plane.
A more detailed description of the CMS apparatus can be found in ref.~\cite{CMS_JINST}.

\subsection{Simulated samples}

Samples of W + jets events are generated with \MADGRAPH5 followed by hadronization and parton showering using the Z2 tune~\cite{Z1} of \PYTHIA6 (version 6.4.25)~\cite{Sjostrand:2006za}.
The \MADGRAPH event generator produces parton-level events  with a W boson and up to four partons in the final state on the basis of matrix element (ME) calculations.
The ME/PS matching scale $\mu$ is taken to be 20\GeV, and the factorization and renormalization scales are set to $q^{2}=M_{\PW}^{2}c^{2} + (\pt^{\PW})^{2}$, where $M_{\PW}$ and $\pt^{\PW}$ are the mass and transverse momentum of the W boson, respectively.

Samples of Z/$\gamma$* + jets and \ttbar events are also simulated with \MADGRAPH5.
Single-top-quark samples are generated with \POWHEG2~\cite{hamilton:2010}.
Samples of WW and WZ events are generated with the {\PYTHIA6} Monte Carlo (MC) event generator.
Contributions of multijet events from QCD interactions are estimated from data, as discussed later.
The inclusive cross sections for simulated processes  are normalized to the next-to-leading-order (NLO), next-to-NLO (NNLO), or next-to-next-to-leading-log (NNLL) order calculations.
Table~\ref{tab:xsec} gives the values of the cross sections of the simulated processes and their theoretical uncertainties.
Theoretical uncertainties have dominant contributions from the PDF uncertainties and the dependence on the renormalization and factorization scales.
All simulations are inclusive in terms of final-state partons. Whenever needed, exclusive samples of W + 0-jet and W + 2-jet events are obtained from the inclusive samples by applying selections on the jet multiplicity at the particle and detector levels.

\begin{table}[htbp]
\centering
\topcaption{\label{tab:xsec} Cross sections of the various processes and their uncertainties. }
\begin{tabular}{l c} \hline
Process & Cross section (pb)\\
\hline
$\PW\rightarrow\mu\nu$ & 10500 $\pm$ 5\% (NNLO)~\cite{wzXsec}\\
$\PW\rightarrow\tau\nu$ & 10500 $\pm$ 5\% (NNLO)~\cite{wzXsec}\\
\cPqt\cPaqt & 160 $\pm$ 7\% (NNLL)~\cite{topXsec} \\
Single top quark & 85 $\pm$ 5\% (NLO)~\cite{topXsec1,topXsec2,topXsec3}\\
Drell--Yan & 3050 $\pm$ 4.3\% (NNLO)~\cite{wzXsec}\\
Diboson (WW + WZ) & 61 $\pm$ 10\% (NLO)~\cite{wwXsec} \\
Multijet & Estimated from data by fitting \\
          & \ETslash distribution in control region\\
\hline
\end{tabular}
\end{table}
The simulated samples are processed and reconstructed in the same manner as the collision data.
The detector response is simulated in detail by using the \GEANTfour package~\cite{Agostinelli:2002hh}.
 The samples include additional interactions per beam crossing (the so-called pileup), which match the corresponding distribution in data.

In addition to these fully simulated samples, various simulations at particle level are compared with the fully corrected DPS-sensitive observables.
\begin{itemize}
\item
\MADGRAPH5 + \PYTHIA8: W + jets events are generated by means of \MADGRAPH5 (as discussed before) followed by hadronization and parton showering using the 4C tune of \PYTHIA8.
The MPI~\cite{Sjostrand:1986ep} are simulated with the \PYTHIA8 event generator.
In order to see the effects of MPI, events are also produced without the MPI contribution by \PYTHIA8.
For the systematic studies, hadronization and parton showering of \MADGRAPH5 events are performed with  \PYTHIA6 tune Z2*~\cite{Z2star}, with and without MPI.
\item
\POWHEG2 + \PYTHIA6 (\HERWIG6): W + 2-jet events are also produced up to NLO accuracy with the \POWHEG2 event generator with the ``Multi-scale improved NLO'' (MiNLO) method~\cite{MINLO}.
The W + 2-jet samples simulated with the \POWHEG2 + MiNLO describe satisfactorily the inclusive W + jet production data as well~\cite{PowhegW2J}.
Hadronization and parton showering is carried out with \PYTHIA6, tune Z2*.
To assess the effect of angular-ordered showering, \HERWIG6 (version 6.520)~\cite{herwig1,herwig2} is also used for the parton showering.
\item
 \PYTHIA8: W + jets events are generated with the 4C tune of the \PYTHIA8 event generator, which produces hard subprocesses with a W boson and either zero or one additional parton in the final
state. It also performs hadronization and parton showering.

\end{itemize}
The MPI model is similar in \PYTHIA6 and \PYTHIA8, with the free parameters tuned to the underlying event data obtained at the LHC.
The key features of the model are:

\begin{itemize}
\item the ratio of the 2$\rightarrow$2 partonic cross section, integrated above a transverse momentum cutoff scale, to the total inelastic pp cross section, which is a measure of the amount of MPI.
A factor with a free parameter, ${{\pt}_0}$, is introduced to regularize an otherwise divergent partonic cross section,

\begin{equation}
\frac{\alpha_{s}^{2}(\pt^{2} + {{\pt}_0}^{2})}{\alpha_{s}^{2}(\pt^{2})} \cdot \frac{\pt^{4}}{(\pt^{2} + {{\pt}_0}^{2})^{2}}\ ,
\end{equation}

with

\begin{equation}
{{\pt}_0} (\sqrt{s}) = {{\pt}_0} (\sqrt{s_{0}})
\left( \frac{\sqrt{s}}{\sqrt{s_{0}}} \right)^{\epsilon} \ .
\end{equation}

Here $\sqrt{s_{0}} = 1.8\TeV$ and $\epsilon$ is a parameter characterizing the energy dependence of ${{\pt}_0}$.
\item A poisson distribution for the number of MPI in an event, with a mean that depends on the overlap of the  matter distribution of the hadrons in impact parameter space.
The impact parameter profile gives a measure of $\sigma_\text{eff}$.
The present model uses the convolution of the matter distributions of the two incoming hadrons as an estimate of the impact parameter profile.
The overlap function is of the form $\re^{-bZ}$, where $b$ is the impact parameter and $Z$ is a free parameter.
\end{itemize}
The MPI model used here~\cite{Sjostrand:2006za} includes parton showers for the MPI processes as well as MPI processes interleaved with initial state radiation.

Events simulated with LO event generators, i.e. \MADGRAPH5, \PYTHIA6, and \PYTHIA8, use the {CTEQ6L}~\cite{Pumplin:2002vw} PDF set, whereas in the NLO event generation with \POWHEG the {CTEQ6M}~\cite{Pumplin:2002vw} PDF set is used.

\subsection{Event selection}
Events were selected online when at least one muon candidate was found.
 A muon candidate consists of a track with hits in the muon system and a transverse momentum greater than a threshold.
 The threshold was increased with increasing instantaneous luminosity in order to keep the rate within the allocated trigger bandwidth for muon triggers.
 The offline selection requires exactly one muon reconstructed in the muon detector and the silicon tracker.
 Muon candidates are required to satisfy identification criteria based on the number of hits in the muon detector and the tracker, their transverse impact parameter with respect to the beam axis, and the goodness of the global fit $\chi^{2}$/(number of degree of freedom)~\cite{CMS_wz} for the tracks in the tracking system and the muon chambers.
 The background from jets misidentified as muons and from semileptonic decays of heavy quarks is suppressed by applying an isolation condition on the muon candidates.
The muon candidate is considered to be isolated if the isolation variable~\cite{UEinDY}, $I$, is smaller than 0.1.

The selected muon is required to have ${\pt} > 35$\GeVc and $\abs{\eta} < 2.1$.
The trigger efficiency for the selected muon is larger than 90\% and the muon selection efficiency is about 95\%~\cite{CMS_wz}.
 The muon candidate is retained only if associated with the primary vertex identified as the signal vertex.
 The selected signal vertex is required to be within $\pm$24\cm of the nominal interaction point along the $z$ direction.
 At least five tracks are required to be associated with the signal vertex, and the transverse displacement of the signal vertex from the beam axis is required to be less than 2\cm.

 Jets and \ETslash are reconstructed with the particle-flow (PF) algorithm~\cite{PF}, which combines information from several sub-detectors.
 The jet reconstruction is based on the anti-\kt clustering algorithm~\cite{antikt,Cacciari:2011ma,Cacciari:2005hq} with a distance parameter of 0.5.
 Jets are required to have ${\pt} > 20$\GeVc and $ \abs{\eta} < 2.0$ to ensure that they are well reconstructed and fall within the  tracker acceptance.
 Jets are required to satisfy identification criteria that eliminate jet candidates originating from noisy channels in the hadron calorimeter~\cite{noise}.
 Jet energy scale (JES) corrections are applied to account for the non-linear response of the calorimeters to the particle energy and other instrumental effects.
 These corrections are based on in-situ measurements using dijet, $\gamma$ + jet, and Z + jet data samples~\cite{jec}.
 Pileup and the underlying event can contribute additional energy to the reconstructed jets.
 The median energy density due to pileup is evaluated in each event and the corresponding energy is subtracted from each jet~\cite{salam:2008}.
 Jets are rejected if they overlap with selected muons within a cone of radius 0.5.
 In order to reject additional jets from pileup interactions, a pileup mitigating variable $\beta$ is utilized, defined as:

\begin{equation}
\beta = \frac{\Sigma  \left(\pt^{\text{signal vertex}} \right)}{\Sigma \left(\pt^\text{all}\right)},
\end{equation}

where $\Sigma (\pt^{\text{signal vertex}} )$ is the sum of the $\pt$ of all charged constituents in a jet  associated with the signal vertex and $\Sigma (\pt^\text{all})$ is the sum of the $\pt$ of all charged constituents in a jet.
Jets are  required to have $\beta > 0.4$ .

The W transverse mass (\MT) is defined as:

\begin{equation}
\MT = \sqrt{2 \cdot \pt^{\mu} \cdot \ETslash \cdot \Big(1 - \cos\big(\Delta\phi[\mu , \ETslash]\big)\Big)} .
\end{equation}

The \ETslash  is defined as the negative vector sum of the transverse momenta of all reconstructed particle candidates in the event, $\ETslash = -\Sigma_i {\ptvec}(i)$.
The reconstructed \ETslash is corrected for the non-compensating nature of the calorimeters and detector misalignment using the procedure described in ref.~\cite{MET}.
This procedure uses all corrected jets which have $\pt> 10$\GeVc and less than 90\% of their energy in the ECAL.
The \ETslash is also corrected for the effect of the azimuthal variation of the tracker acceptance and the calorimeter alignment.
The correction factor is calculated as a function of the number of reconstructed vertices and also as a function of $\Sigma \ET$, where $\Sigma \ET$ is the total transverse energy measured in the calorimeter.
The angle $\Delta\phi[\mu, \ETslash]$ is measured between the muon $\mu$ and the \ETslash direction in the azimuthal plane.

Events are required to have exactly one muon with $\pt> 35$\GeVc, $\abs{\eta} < 2.1$, and $\ETslash > 30$\GeV.
The transverse mass is required to be greater than 50\GeVcc.
Selected events are required to have exactly two jets with $\pt> $ 20\GeVc and $\abs{\eta} < $ 2.0.
The criteria used in the selection are summarized in Table~\ref{tab:sel}.

\begin{table}[htbp]
\centering
\topcaption{\label{tab:sel} Summary of the W + 2-jet event selection and reconstruction criteria at the detector level. }
\begin{tabular}{l | c} \hline
W $\rightarrow~\mu\nu$ selection & Jet selection\\
\hline
Single-muon trigger &   Anti-\kt PF jet with $R = 0.5$\\
Muon ID and isolation & ${\pt} > 20$\GeVc, $\abs{\eta} < 2.0$\\
Exactly one muon $\pt> 35$\GeVc, $\abs{\eta} < 2.1$ & $\beta > 0.4$\\
$\ETslash > 30\GeVc$ &   $\Delta R_{{\rj,}\mu} > 0.5$\\
W-boson transverse mass $> 50\GeVcc$ &       exactly two jets           \\
\hline
\end{tabular}
\end{table}

\begin{figure}[htbp]
\centering
\includegraphics[width=0.48\textwidth]{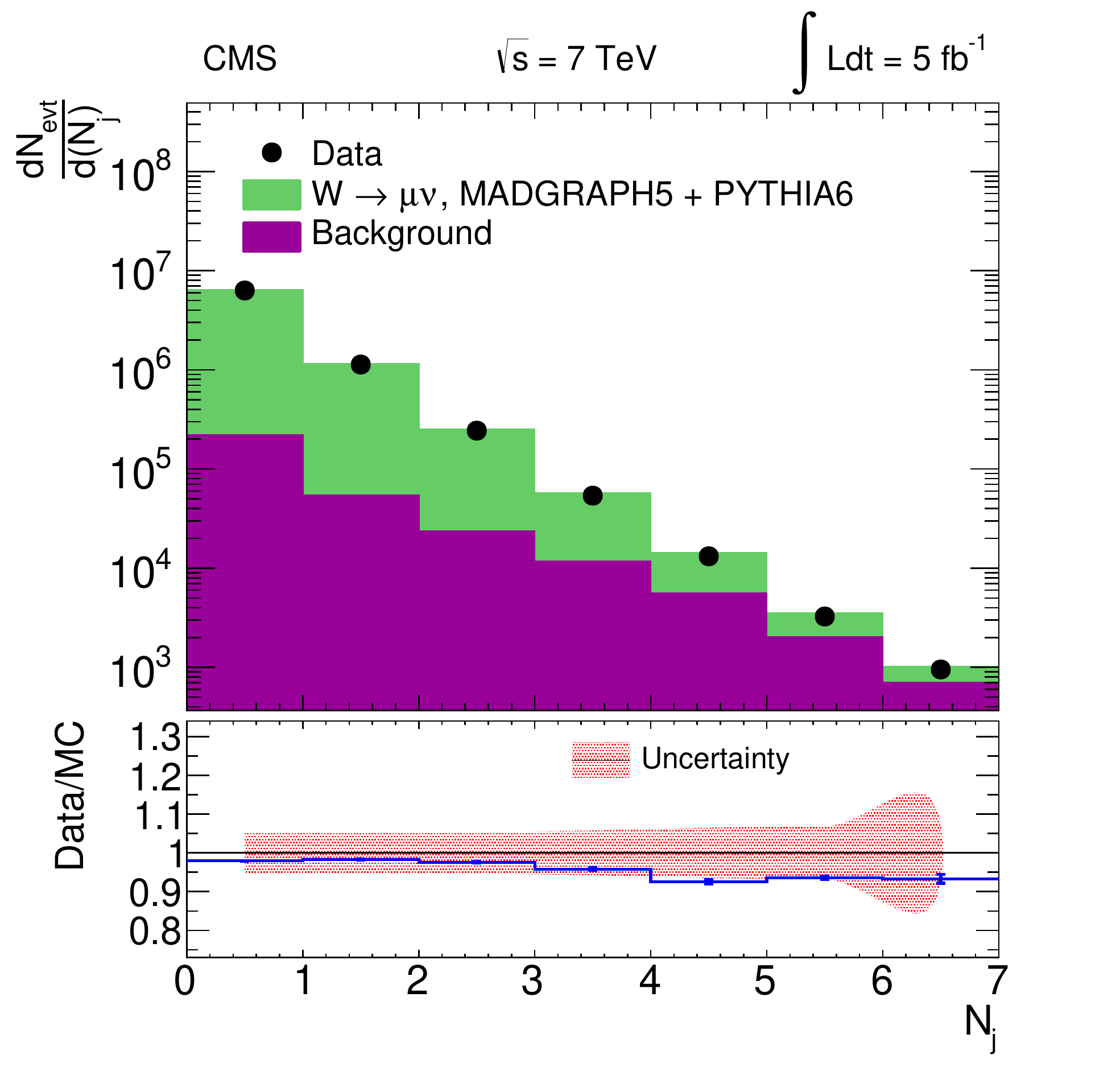}
\includegraphics[width=0.48\textwidth]{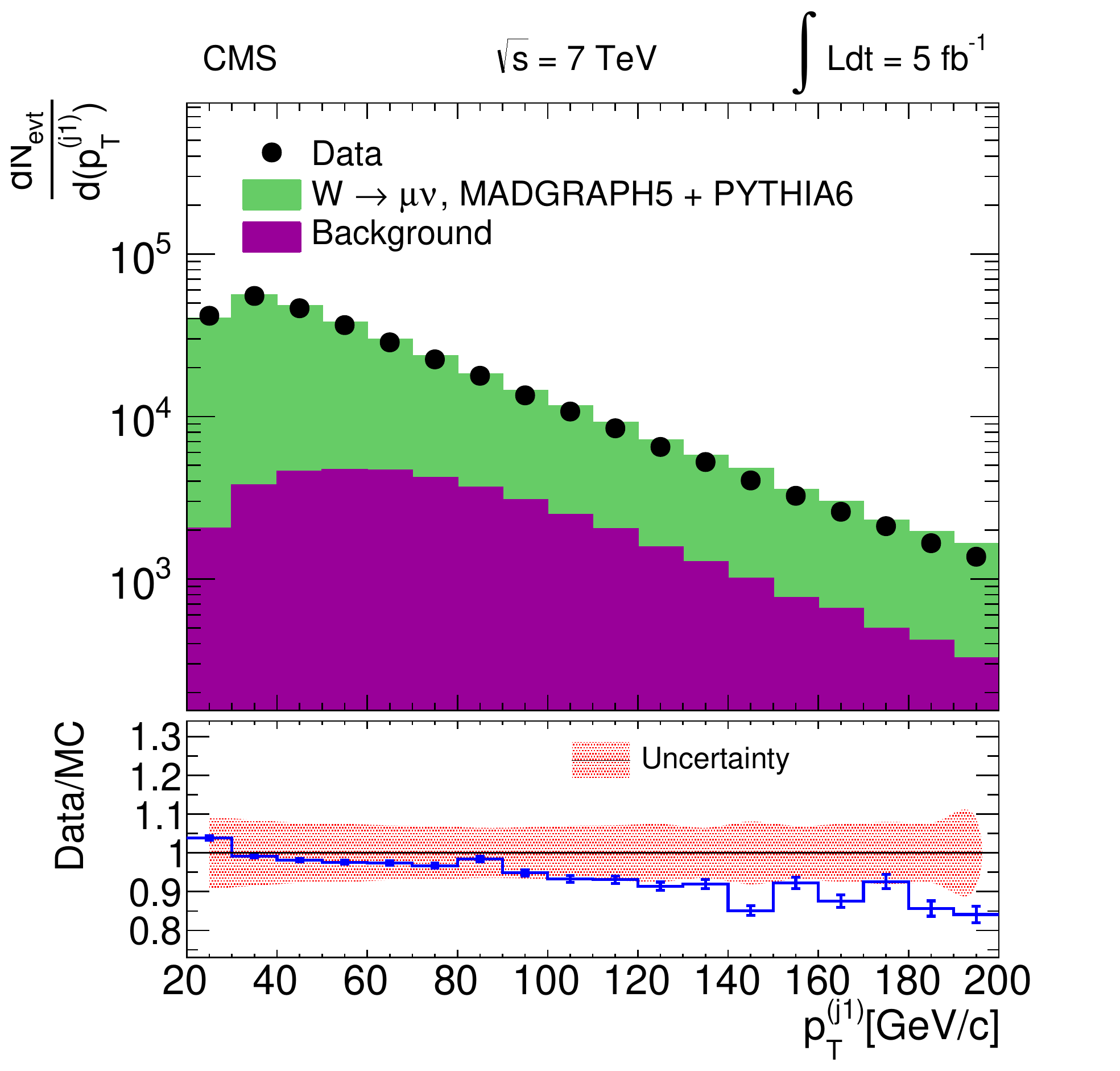}
\includegraphics[width=0.48\textwidth]{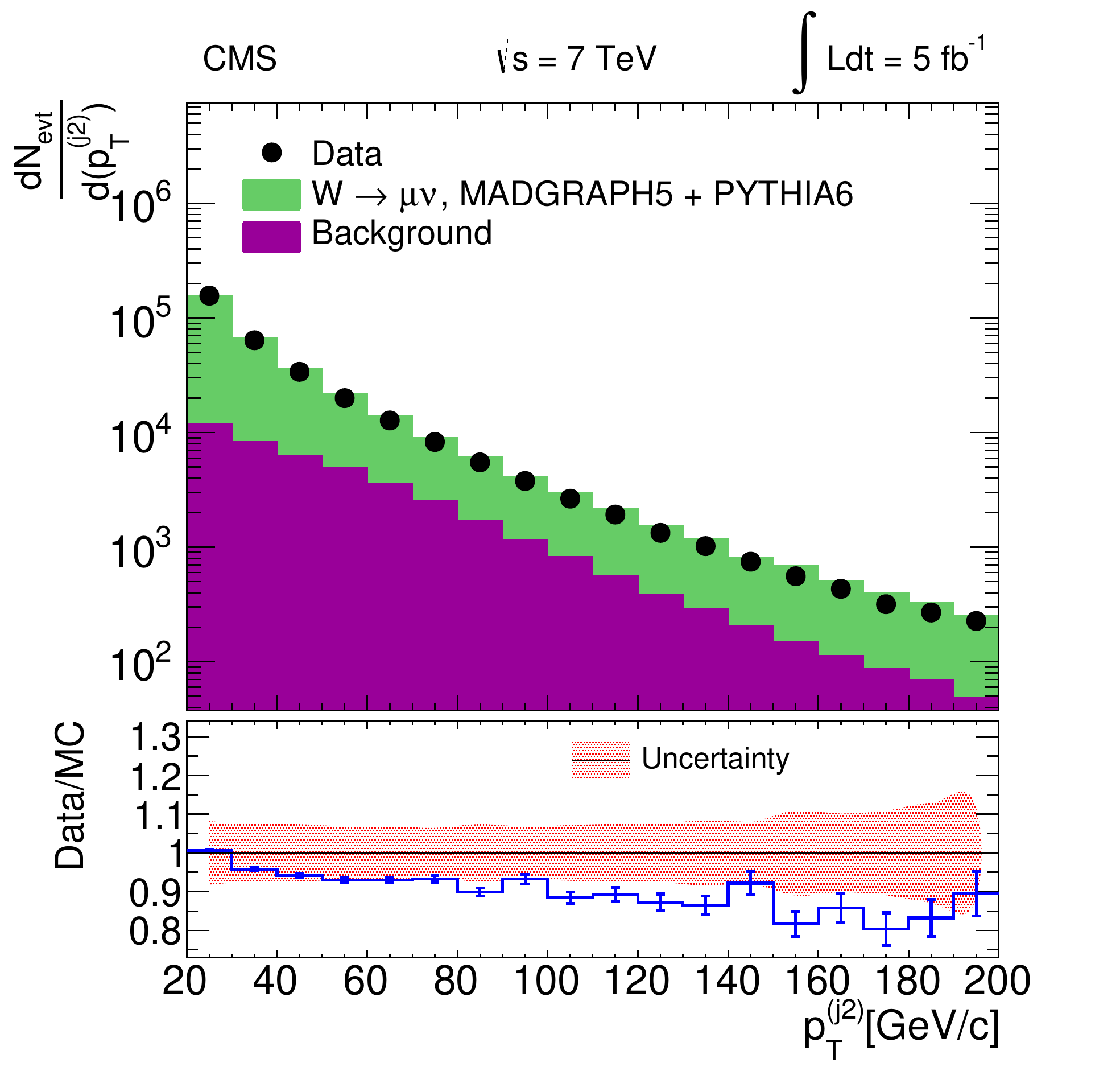}
\includegraphics[width=0.48\textwidth]{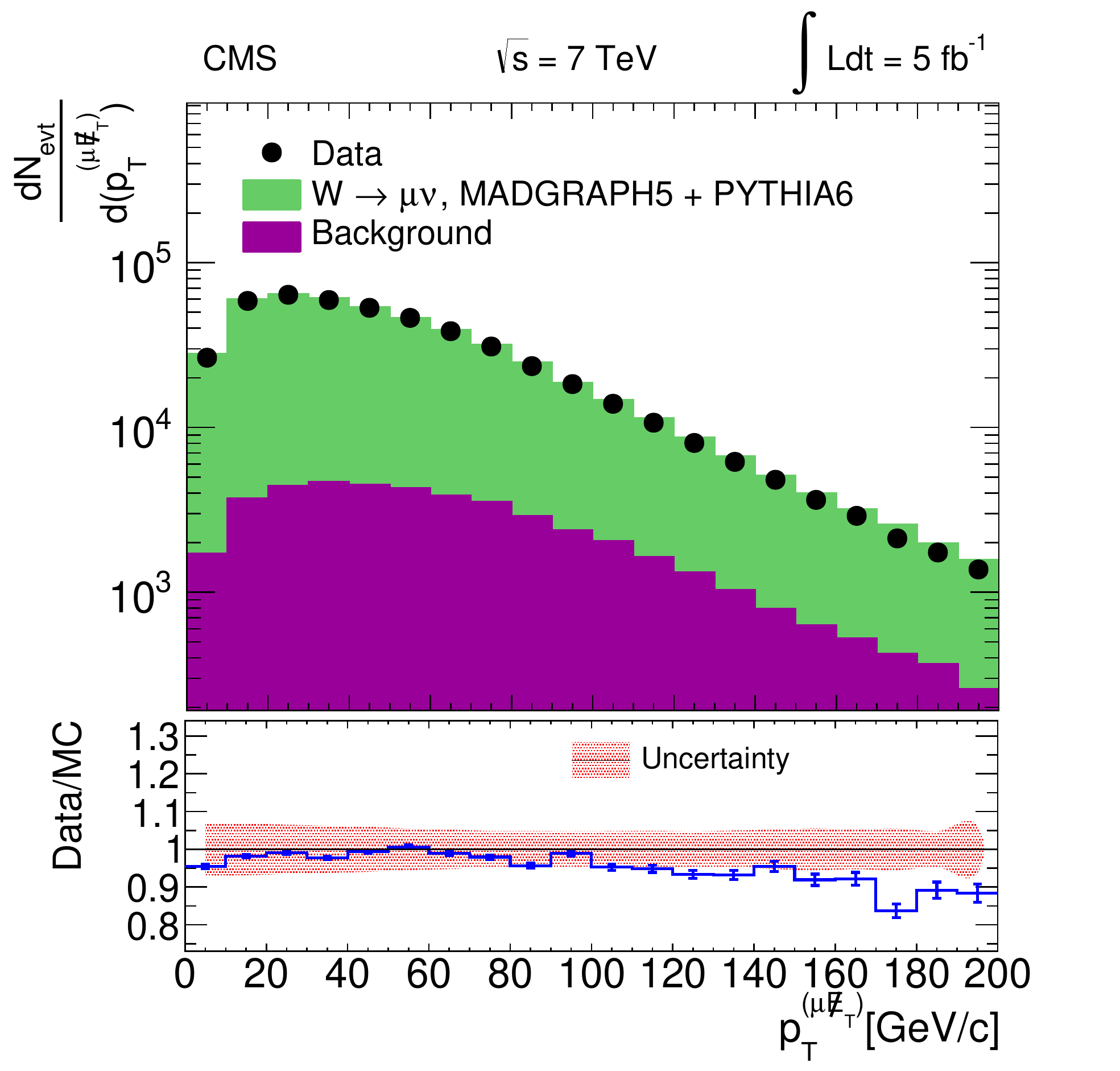}
\caption{Detector-level comparison of data with MC simulations for the multiplicity (top left) of jets (N$_{\rj}$) with ${\pt} > 20$\GeVc and $\abs{\eta} < $ 2.0.
Data and simulations for the sample with exactly two jets are plotted as a function of the $\pt$ of the leading (top right) and subleading (lower left) jets, as well as of the magnitude of the vector sum of the muon $\pt$ and \ETslash (lower right).
The background distribution represents the sum of the contributions of Drell--Yan, $\PW\rightarrow~\tau\nu$, diboson, multijet, \cPqt\cPaqt, and single-top-quark processes.
The bottom panels show the ratio of the data and simulated distributions.
The band shows the total uncertainty, with the contributions of the jet energy scale uncertainty and the statistical uncertainties of the MC samples added in quadrature.
The error bars on the ratio histogram represent the statistical uncertainty of the data and the simulated samples added in quadrature.
}
\label{fig:val_rec1}
\end{figure}

\begin{figure}[htbp]
\centering
\includegraphics[width=0.48\textwidth]{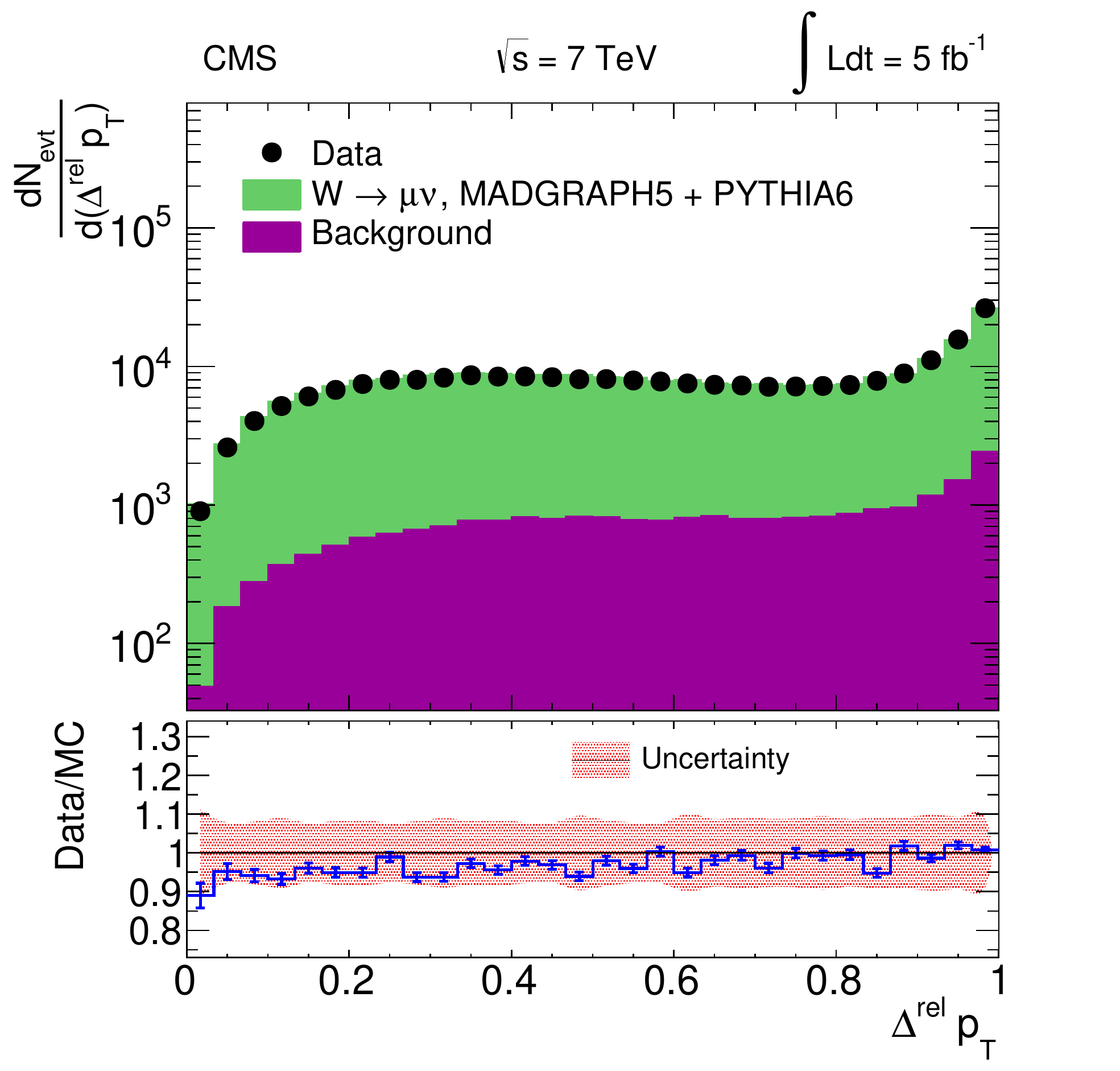}
\includegraphics[width=0.48\textwidth]{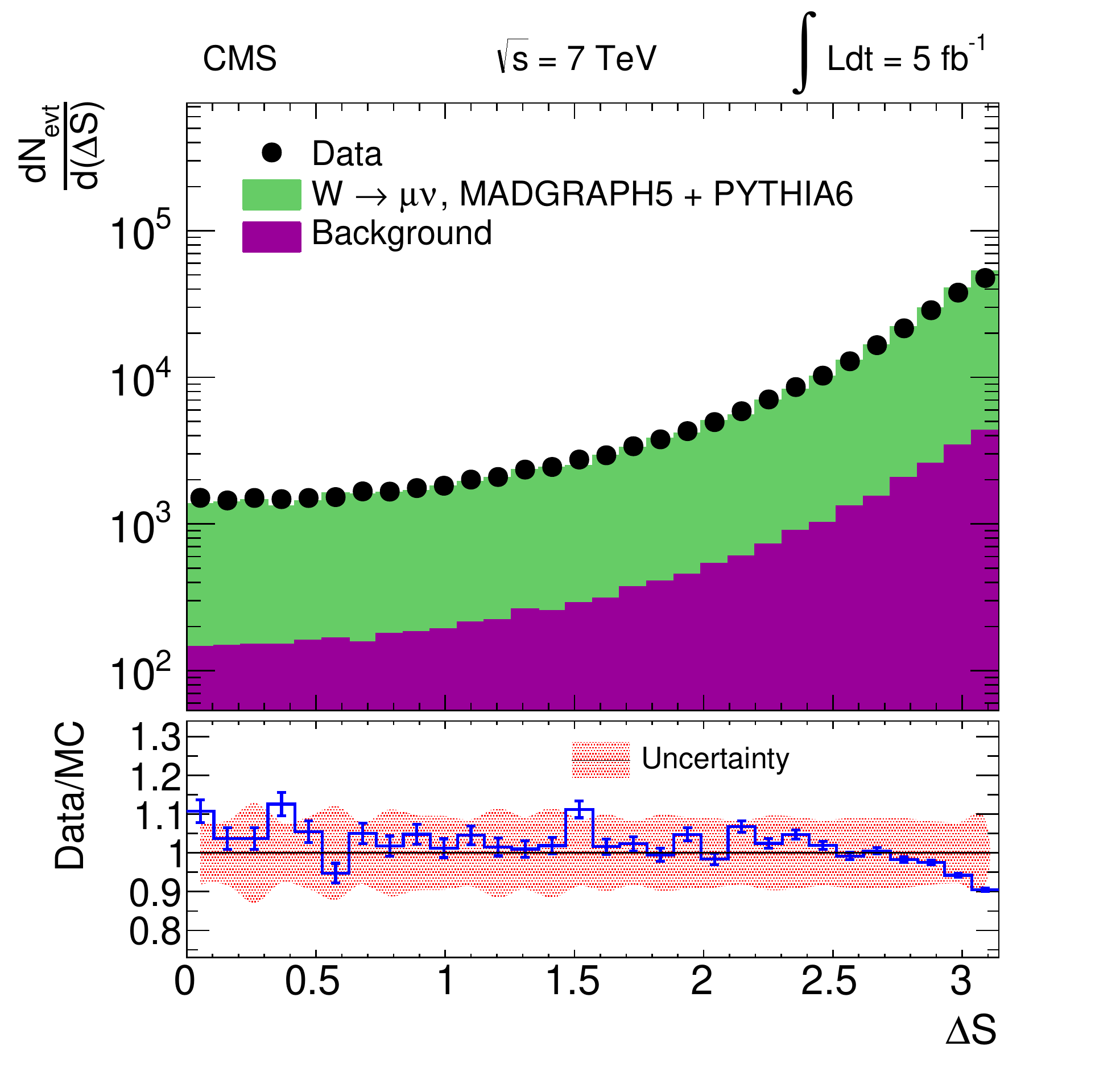}
 \caption{Comparison of data with MC simulations at detector level for the DPS-sensitive observables $\Delta^\text{rel}~ \pt$ (left), and $\DS$ (right).
The background distribution represents the sum of the contributions of Drell--Yan, W $\rightarrow~\tau\nu$, diboson, multijet, \cPqt\cPaqt, and single-top-quark processes.
The bottom panels show the ratio of the data and simulated distributions.
The band shows the total uncertainty, with the contributions of the jet energy scale uncertainty and the statistical uncertainties of the MC samples added in quadrature.
The error bars on the ratio histogram represent the statistical uncertainty of the data and the simulated samples added in quadrature.}
\label{fig:rec_val}
\end{figure}

The kinematic distributions of the jets in the selected events are reproduced by the MC simulations as shown in Fig.~\ref{fig:val_rec1}.
Figure~\ref{fig:rec_val} shows the comparison of data with MC simulations for $\Delta^\text{rel}~ \pt$~(left) and $\DS$~(right) at the detector level.
Data and MC simulation are in good agreement except for a 10--20\% difference at large $\pt$.
This difference is not a concern for the present analysis because the contribution of events having large $\pt$ jet is very small, e.g., only 3\% of the W + 2-jet events have a jet $\pt$ larger than 100\GeVc, where the description starts to deviate from the data.
There is a small level of background contamination in the selected W + 2-jet samples.
The dominant background contribution comes from top-quark production (single top-quark and pair production) and Drell--Yan processes.
The contribution of the multijet background is less than 0.5\%.
This contribution is estimated by defining  a control region with the requirement of a non-isolated muon, with $I > 0.1$.
A template of the \ETslash distribution for the multijet background is constructed by using events in this control region.
 This template is then used to estimate the contribution of the multijet background by fitting the \ETslash distribution in the signal region, with $I < 0.1$.
Table~\ref{tab:evt_sum} summarizes  the expected number of events for the processes listed in Table~\ref{tab:xsec}, for an integrated luminosity of 5\fbinv.
The total contribution from the background events is about 10\% but the effect on the shape of the DPS-sensitive observables is less than 1\%.

\begin{table}[htbp]
\centering
\topcaption{\label{tab:evt_sum}Expected yields for various processes for 5\fbinv and observed number of events in the data.
The top production background is the sum of the single-top-quark and \cPqt\cPaqt\ processes.
The estimated event yields from the simulated samples include uncertainties in the respective cross sections.
}
\begin{tabular}{l c} \hline
Process & Number of events \\
\hline
$\PW\rightarrow\mu\nu$& (2.3 $\pm$ 0.12) $\times$ 10$^{5}$ \\
$\PW\rightarrow\tau\nu$& (3.7 $\pm$ 0.20) $\times$ 10$^{3}$ \\
Top quark& (9.4 $\pm$ 0.69) $\times$ 10$^{3}$ \\
Drell--Yan& (5.3 $\pm$ 0.26) $\times$ 10$^{3}$\\
Diboson & (2.6 $\pm$ 0.26) $\times$ 10$^{3}$\\
Multijet & (1.1 $\pm$ 0.34) $\times$ 10$^{3}$\\
\hline
Total expected events & (2.5 $\pm$ 0.14) $\times$ 10$^{5}$ \\
\hline
Data & (2.4 $\pm$ 0.0049) $\times$ 10$^{5}$\\
\hline
\end{tabular}
\end{table}

\section{Unfolding and comparison with simulations}
\label{sec:correction}
The sample of W + 2-jet events is selected as discussed in the previous section.
The contributions of all backgrounds are subtracted from the data distributions before unfolding.

The distributions of the DPS-sensitive observables for the selected events are corrected for selection efficiencies and detector effects.
The trigger efficiency does not bias the shape of the DPS-sensitive observables; this was checked by comparing simulated samples with and without the trigger requirement.
The selected events are mainly SPS events and the sample contains only a few percent of DPS events.
However, in the DPS-sensitive region, at low $\Delta ^\text{rel}~{ \pt}$ and $\DS$, the DPS contribution is relatively large (a few tens of percent).
Thus, the shape of the distribution of $\Delta ^\text{rel}~{ \pt}$ and $\DS$ is more important than the absolute normalization in the extraction of the DPS fraction.
Therefore, the unfolding and the systematic studies are carried out for the shapes of the $\Delta ^\text{rel}~{\pt}$ and $\DS$ distributions.
The measured distributions are unfolded to the level of stable particles (lifetime $c\tau > 10$ mm) within the phase space given in Table~\ref{tab:sel1}.

\begin{table}[thb]
\centering
\topcaption{\label{tab:sel1} Phase space definition for the visible cross section at the particle level. }
\begin{tabular}{l c} \hline
1 $\mu$ : ${\pt} > 35 $\GeVc and $\abs{\eta} < 2.1$ \\
$\ETslash > 30$\GeV and \MT $ > 50$\GeVcc \\
Exactly 2 jets : ${\pt} > 20$\GeVc and $\abs{\eta} < 2.0$   \\
\hline
\end{tabular}
\end{table}

Unfolding is performed with an iterative Bayesian method~\cite{unfold} that properly takes into account bin-to-bin migrations.
A response matrix is created with simulated events produced with the \MADGRAPH5 + \PYTHIA6 Monte Carlo event generator.
The unfolding is cross-checked by using the singular value decomposition (SVD)~\cite{svd} approach.
Iterative Bayesian and SVD approaches give consistent results within uncertainties.
Various systematic effects are considered and are listed below:
\begin{itemize}
\item
\textbf{Model dependence:}
The sensitivity to the model dependence of the simulations used for the unfolding is estimated by comparing the results unfolded with different MCs.
 The main effect is due to the simulation of MPI, and is estimated by comparing the detector-level distributions of $\Delta ^\text{rel}~{\pt}$ and $\DS$ unfolded by using \MADGRAPH5 + \PYTHIA6, with and without MPI.
 The effect of removing MPI is about 3--4\%, independent of $\Delta ^\text{rel}~{\pt}$ and $\DS$.
 This is taken as an estimate of the systematic uncertainty due to the model dependence of the simulations.
\item
\textbf{Background subtraction:} The contribution from various backgrounds is estimated with simulated samples that are subtracted from data before applying any corrections.
In order to estimate the systematic uncertainties, the cross sections of the background processes are varied within their uncertainties.
The shape of the background distribution is affected by the jet energy scale and \ETslash uncertainties.
The total effect of all these uncertainties on the final distribution is less than 0.5\%.
\item
\textbf{Jet energy scale (JES):} The four momentum of each jet is varied by the JES uncertainty.
This variation gives a systematic bias of 1--3\%.
\item
\textbf{Jet energy resolution (JER):} The JER is different between data and simulation by 3--8\% for $\abs{\eta} < 2.0$.
A variation of the JER by this amount in the simulation affects the distribution by less than 1\%.
\item
\textbf{Resolution of \ETslash:}  The \ETslash resolution differs in data and simulation~\cite{MET}; this affects the $\DS$ distribution by at most 3.7\%.
The effect on the $\Delta^\text{rel}~ \pt$ distribution is less than 1\%.
\item
\textbf{Pileup:} In order to take into account the uncertainty in the luminosity measurement~\cite{cmsLumi} and the total inelastic cross section, an uncertainty of 5\% is assigned to the mean value of the pileup distribution.
This uncertainty affects the $\DS$ distribution by at most 3.7\%, whereas the effect on the $\Delta^\text{rel}~\pt$ distribution is less than 1\%.
\end{itemize}
Table~\ref{tab:sys1} summarizes the systematic uncertainties for the $\Delta ^\text{rel}~{\pt}$ and $\DS$ distributions.
The absolute cross section of W + 2-jet events is not important for the extraction of the DPS contribution.
However, for completeness the W+ 2-jet production cross section is also corrected to the particle level.
The total cross section for the W + 2-jet production (including the DPS contribution), within the region defined in Table~\ref{tab:sel1}, is measured to be  $53.4\pm 0.1\stat \pm 7.6\syst\unit{pb}$.
 This is consistent with the particle-level prediction by \MADGRAPH5 + \PYTHIA8, scaled by the ratio of the NNLO and LO cross section for inclusive W production,  yielding $55.6\pm2.8\unit{pb}$.
Various systematic effects, arising from the sources discussed above, are also evaluated for the total W + 2-jet cross section.
In addition to these, the cross section has a systematic uncertainty of 2.2\% due to the luminosity measurement~\cite{cmsLumi}.
There is a further uncertainty of 1\% in the trigger efficiency and 2\% in the muon identification and selection efficiencies~\cite{CMS_wz}.
A summary of the various systematic uncertainties for the W + 2-jet cross section is given in the last column of Table~\ref{tab:sys1}.

\begin{table}[t]
\centering
\topcaption{\label{tab:sys1} Summary of the systematic uncertainties (in \%) for different observables.
Uncertainties in integrated luminosity, muon identification (ID), and trigger efficiency only affect the W + 2-jet cross section measurement.
}
\begin{tabular}{l c c c} \hline
Source &  $\Delta ^\text{rel} \pt$ & $\DS$ & Cross section\\
\hline
Model dependence & $\leq$ 3.2 & $\leq$ 3.9 & 11\\
Background normalization & $\leq$ 0.2 & $\leq$ 0.3 & 1.0\\
JES & $\leq$ 1.4 & $\leq$ 2.9 & 7.4\\
JER & $\leq$ 0.5 & $\leq$ 0.7 & 1.3\\
\ETslash scale  & $\leq$ 0.5 & $\leq$ 3.7 & 3.3\\
Pileup & $\leq$ 0.8 & $\leq$ 3.7 & 2.3\\
Muon ID and trigger & ---  & --- & 2.2\\
Luminosity & --- & --- & 2.2\\
\hline
Total  & $\leq$ 3.7 & $\leq$ 7.2 & 14\\
\hline
\end{tabular}
\end{table}

\begin{figure}[htbp]
\centering
\includegraphics[width=0.48\textwidth]{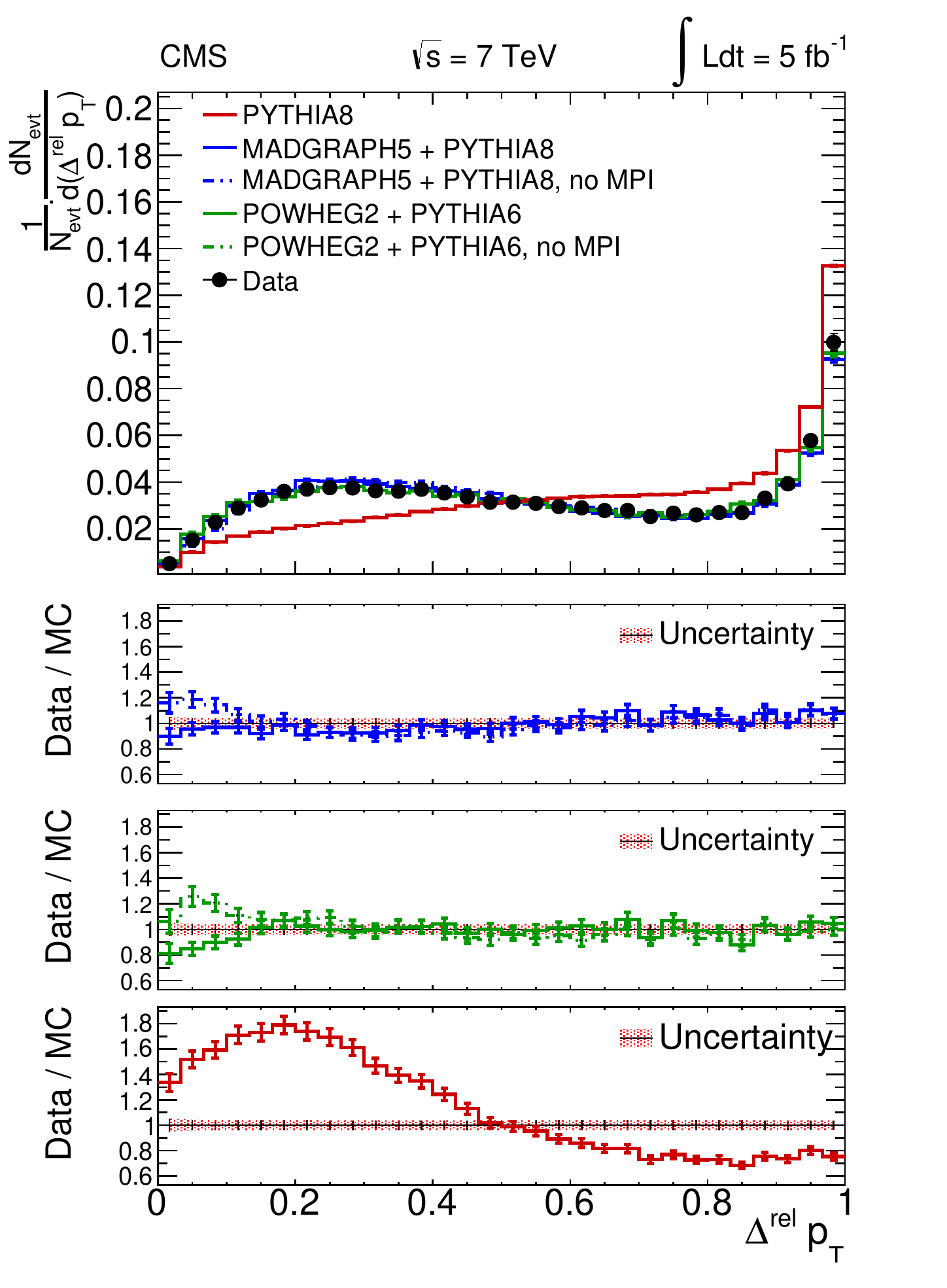} \includegraphics[width=0.48\textwidth]{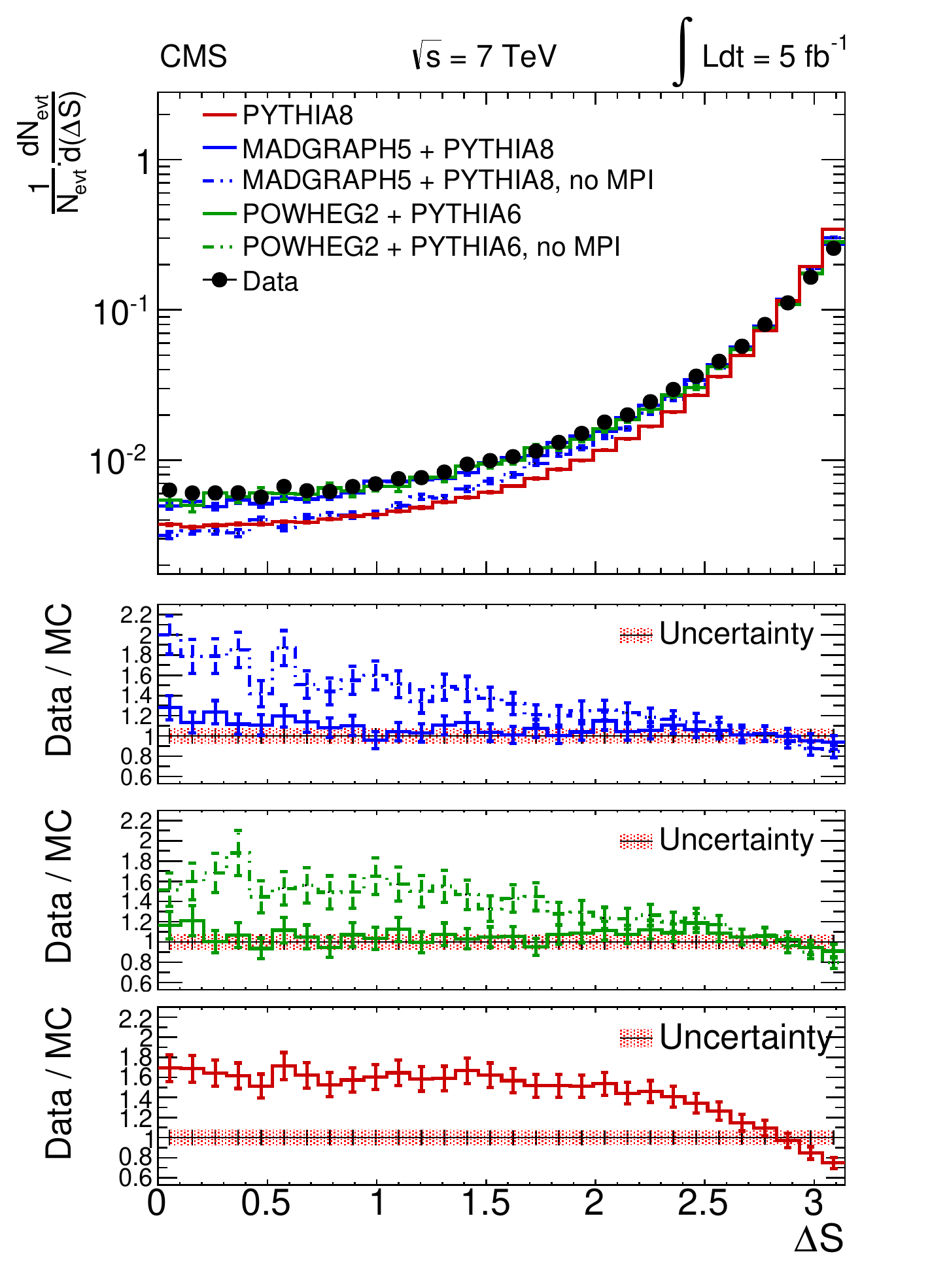}
\caption{Fully corrected data distributions, normalized to unity, for the DPS-sensitive observables $\Delta^\text{rel}~\pt$ (left) and $\DS$ (right).
The second panel in both plots shows the ratio of data over \MADGRAPH5 + \PYTHIA8 with and without MPI, whereas in the third panel the ratio with \POWHEG2 + \PYTHIA6 is shown.
The ratio of the data and \PYTHIA8 is shown in the fourth panel of both plots.
The band represents the total uncertainty of the data (cf. Table 5).
}
\label{fig:cor_norm}
\end{figure}

A comparison of various simulations for inclusive W production with the corrected distributions is shown in Fig.~\ref{fig:cor_norm}.
The $\Delta ^\text{rel}~{\pt}$ and $\DS$ distributions are properly described by \MADGRAPH5 + \PYTHIA8.
\MADGRAPH5, with hadronization and parton showering carried out with \PYTHIA6, also describes the measurements well.
The NLO predictions for W + 2-jet production obtained with \POWHEG2 + \PYTHIA6 also satisfactorily describe the data.
Measurements are also well reproduced by \POWHEG2, with hadronization and parton showering carried out with \HERWIG6.

The \PYTHIA8 simulation underestimates the measurements by a factor of 1.5--2.0.
This discrepancy is due to the fact that \PYTHIA8 generates only 2$\to$1 and 2$\to$2 processes and most of the additional jets are produced during parton showering, and have a softer \pt spectrum than that measured in data.
The difference is mainly in the DPS-sensitive region.
Therefore, event generators used to define SPS backgrounds must include a proper implementation of additional hard radiation.
 If it is not included, the effect of missing hard radiation might be interpreted as a DPS contribution.
Without MPI, the LO and NLO predictions from \MADGRAPH5 + \PYTHIA8 and \POWHEG2 + \PYTHIA6 are unable to describe the data shown in Fig.~\ref{fig:cor_norm}.
The importance of including MPI, for both LO and NLO simulations in the description of W + 2-jet events, is conclusively shown by the comparisons of data with simulations with and without MPI.
In the following sections the contribution of hard MPI is extracted.

\section{Strategy for the extraction of the DPS fraction}
\label{sec:sigbkg}

The fraction of W + 2-jet events produced by DPS is extracted by performing a template fit to the fully corrected distributions of $\Delta ^\text{rel}~{\pt}$ and $\DS$ using a binned likelihood method.
Here, the strategy to extract the DPS fraction is discussed, including the definition of the signal and background templates and the corresponding systematic uncertainties.

\subsection{DPS signal template}
In this analysis, DPS events are required to have one W boson with zero jets from the first interaction and two jets from an independent second interaction, where the jets are required to have ${\pt} > 20$\GeVc and $\abs{\eta} < 2.0$.
In present analysis, the two interactions are assumed to be independent of each other and a DPS template is produced by randomly mixing W and dijet events.
The DPS template is produced by mixing dijet events simulated with \PYTHIA8 and W + 0-jet events.
 These W + 0-jet events are selected as a subsample of the W + jets inclusive sample simulated with \MADGRAPH5 + \PYTHIA8. 
The DPS templates produced with simulated events are in good agreement with the templates obtained by mixing W + 0-jet and dijet events in the data.

\begin{figure}[htbp]
\centering
\includegraphics[width=0.48\textwidth]{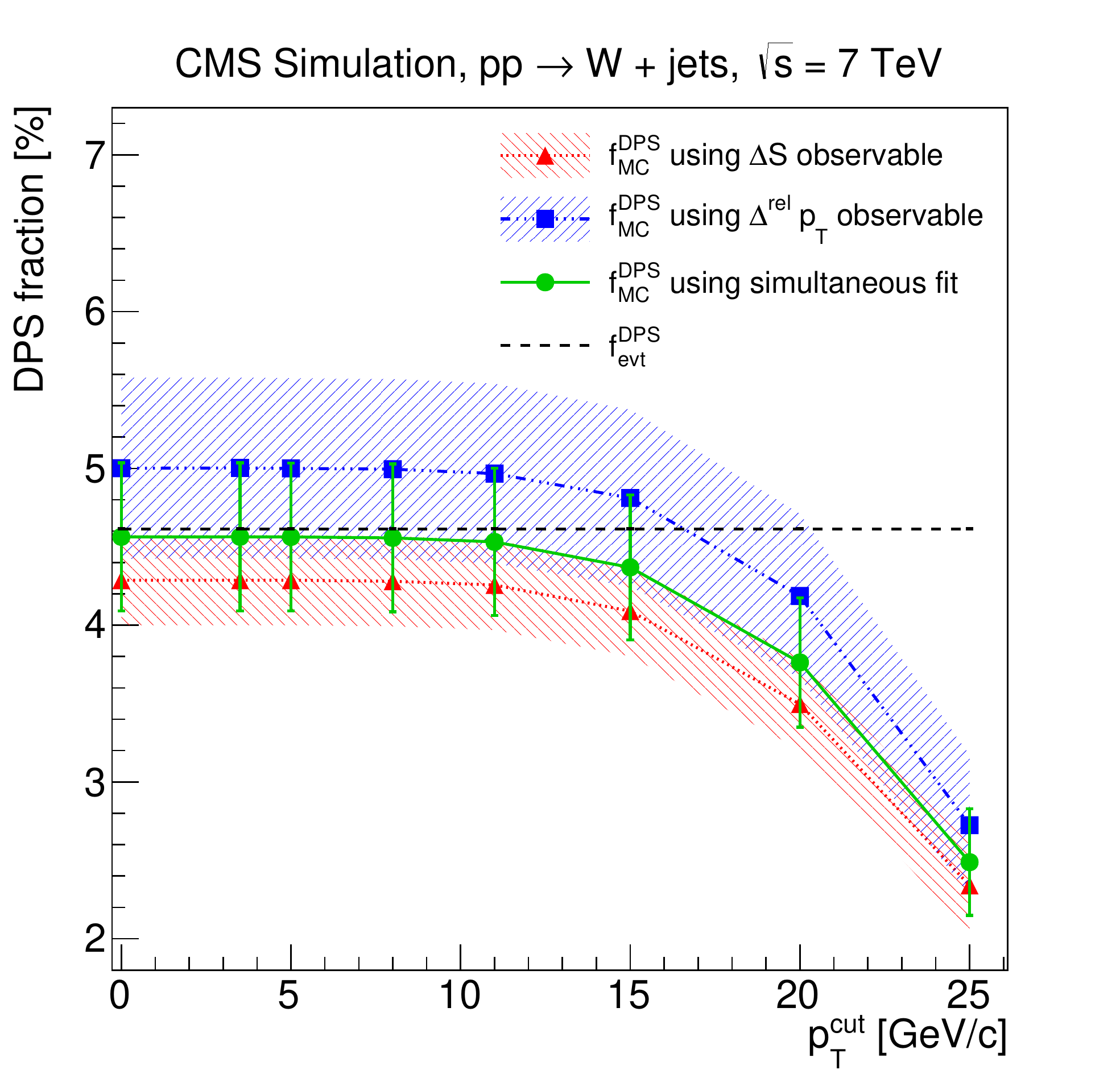}
\caption{The extracted value of the DPS fraction in W + jets events, simulated with \MADGRAPH5 + \PYTHIA8,  using different background templates obtained by varying the transverse momentum cutoff (${\pt^\text{cut}}$) for the second hard interaction.
The DPS fractions obtained by performing both simultaneous and individual fits to the $\Delta ^\text{rel}~{\pt}$ and $\DS$ observables are shown. The DPS fraction, ${f_\mathrm{DPS}^\text{evt}}$, for the simulated  W + jets events is shown by a dashed black line.
The error bars/bands represent the statistical uncertainty added in quadrature to the systematic uncertainty of the DPS template (as discussed in section~\ref{sec:fraction}).
}
\label{fig:ptmax}
\end{figure}

\subsection{Background template}
In order to construct the background template, inclusive W + jets events are used which are produced with \MADGRAPH5 followed by hadronization and parton showering with \PYTHIA8.
The background template cannot be obtained by simply switching off MPI, because only the MPI events that satisfy the DPS signal definition as discussed above need to be excluded.
This requires the tagging of the MPI partons and is achieved, in the case of parton showering with \PYTHIA8, by using the status codes 31--39, which are assigned to partons produced from MPI.
The background template is constructed by taking inclusive W + jets events that contain exactly two jets with  ${\pt} > 20$\GeVc and $\abs{\eta} < 2.0$ and by removing those with two MPI tagged partons with $\abs{\eta} < 2.0$.
It has been argued~\cite{atlas} that the templates for signal and background must be disjoint and that an additional cutoff on the transverse momentum, ${\pt^\text{cut}}$, must be applied to the tagged MPI partons.
By applying the ${\pt^\text{cut}}$, events having MPI partons with $\pt> \pt^\text{cut}$ are removed from the background template.
The effect of applying such a cutoff is studied by comparing at simulation level the fitted and true DPS fraction in the \MADGRAPH5 + \PYTHIA8 sample as a function of ${\pt^\text{cut}}$ (Fig.~\ref{fig:ptmax}).
We observe that, for ${\pt^\text{cut}} <$ 12\GeVc, the fitted DPS fraction does not depend on ${\pt^\text{cut}}$ and therefore we do not apply any cut on $\pt$ of the MPI partons.
The true DPS fraction in \MADGRAPH5 + \PYTHIA8 simulation is obtained by counting events containing a W boson and exactly two jets within the acceptance at particle level.
The two MPI-tagged partons must also be within the acceptance ($\abs{\eta} < 2.0$) and there must be no parton with $\abs{\eta} < 2.0$ from the first interaction.
 This fraction, ${f_\mathrm{DPS}^\text{evt}}$, is determined to be

\begin{equation}
{f_\mathrm{DPS}^\text{evt}} =  0.046 \pm 0.001\stat.
\end{equation}

Because of the different sensitivities of $\Delta ^\text{rel}~{\pt}$ and $\DS$ to DPS, the DPS fraction obtained by fitting only the $\DS$ observable underestimates ${f_\mathrm{DPS}^\text{evt}}$, whereas the fitted result with $\Delta ^\text{rel}~{\pt}$ overestimates the ${\rm f_\mathrm{DPS}^\text{evt}}$.
 However, the fitted DPS fractions are compatible with each other within their systematic uncertainties.
 In these simulation studies, the main source of systematic bias is the model dependence of the signal templates (a detailed discussion is given in section~\ref{sec:fraction}) used for the DPS extraction.
If a simultaneous fit of the $\Delta ^\text{rel}~{\pt}$ and $\DS$ observables is performed, the fitted fraction is consistent with ${f_\mathrm{DPS}^\text{evt}}$.

The two observables $\Delta ^\text{rel}~{\pt}$ and $\DS$ are not correlated for signal events; conversely a 40\% correlation is present for the background events.
The simultaneous fit does not take into account the correlation between the two observables.
The possible effect of the correlation is studied with a simulated sample, by extracting the DPS fraction from a fit to the 2-dimensional distribution of $\Delta ^\text{rel}~{\pt}$ and $\DS$.
The result differs by 4\% from that obtained by simultaneously fitting the two one-dimensional distributions.
This effect of the correlation on the DPS fraction is small as compared to the total systematic uncertainty of 26\% (as discussed in section~\ref{sec:fraction}).

\begin{figure}[htbp]
\centering
\includegraphics[width=0.48\textwidth]{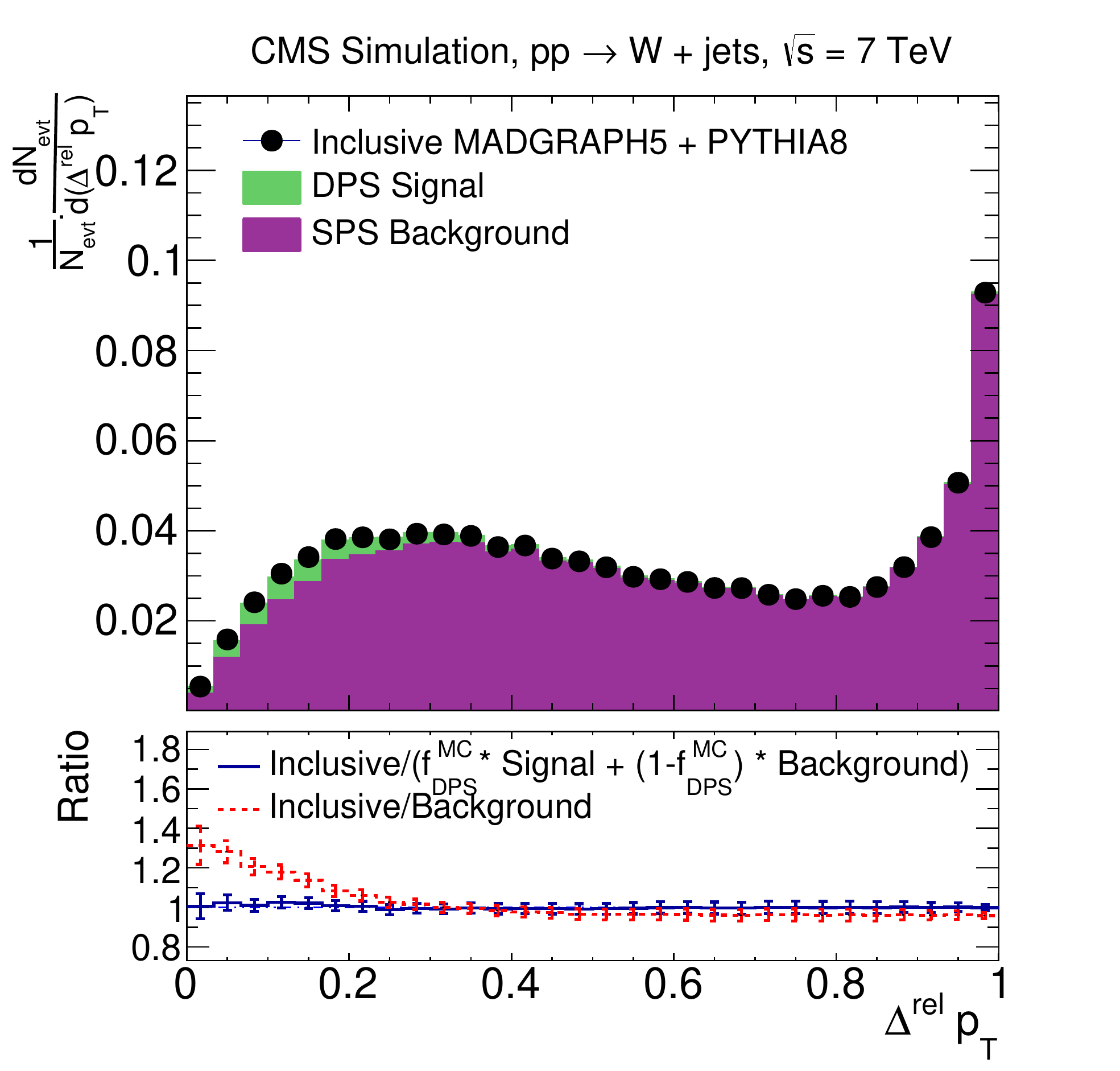} \includegraphics[width=0.48\textwidth]{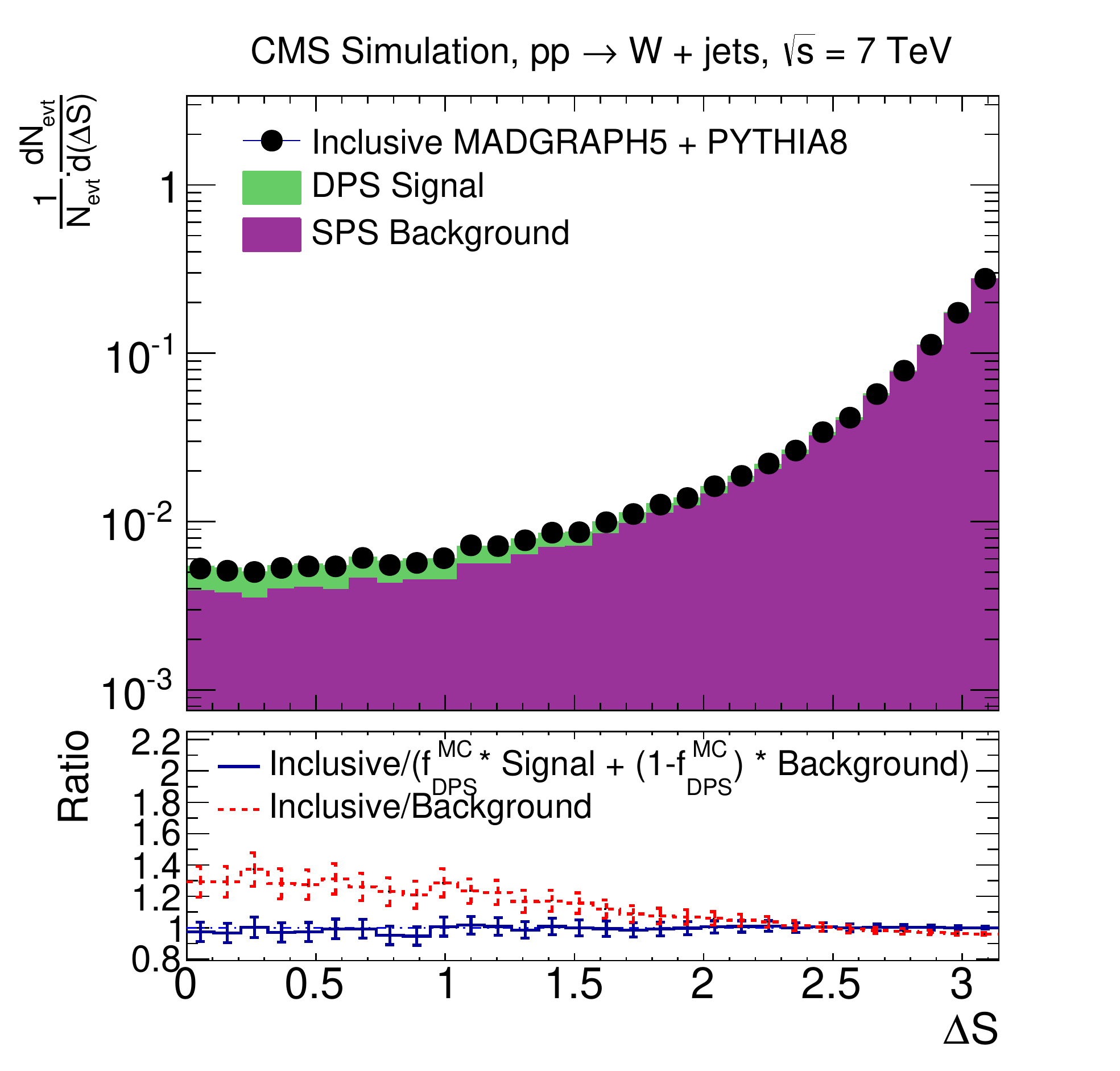}
\caption{Fit results for the DPS-sensitive observables $\Delta ^\text{rel}~{\pt}$ (left) and $\DS$ (right)
 using signal and background templates.
The distributions of the simulated W + 2-jet events are fitted with signal and background templates.
The bottom panels show the ratio of the distributions to the fit results.
Here, the term ``inclusive'' means the simulation also includes the DPS contribution.
\label{fig:fit_mc}}
\end{figure}

Figure~\ref{fig:fit_mc} shows the results of fitting the $\Delta ^\text{rel}~{\pt}$ and $\DS$ observables for simulated W + 2-jet events with signal and background templates.
The extracted DPS fraction in the simulated events is

\begin{equation}
{f_\mathrm{DPS}^{\rm MC}} = 0.045 \pm 0.002~({\rm stat.}),
\end{equation}

which is consistent with the $f_\mathrm{DPS}^\text{evt}$ value predicted by the default MPI model present in the \MADGRAPH5 + \PYTHIA8 simulation.
This closure test also works well when fitting pseudo-data obtained by mixing simulated signal and background events in different proportions.

To summarize, we perform a simultaneous fit of the $\Delta ^\text{rel}~{\pt}$ and $\DS$ observables to utilize their different sensitivities and to reduce the uncertainties.
The signal template is obtained by randomly mixing independently produced W and dijet events, whereas the background template is produced from  the W + 2-jet sample simulated with \MADGRAPH5 + \PYTHIA8, in which events with MPI-tagged partons within the acceptance ($ \abs{\eta} < 2.0$) are removed.

\section{The DPS fraction in data}
\label{sec:fraction}

The $\Delta ^\text{rel}~{\pt}$ and $\DS$  distributions are simultaneously fitted by using the signal and background templates defined in section~\ref{sec:sigbkg}.
The fitted value of the DPS fraction (${f_\mathrm{DPS}}$) is:

\begin{equation}
 {f_\mathrm{DPS}} = 0.055 \pm 0.002\stat \pm 0.014\syst.
\end{equation}

The MC predictions using the fit results are shown in Fig.~\ref{fig:fit_data} and compared to data.

The following sources of systematic uncertainties are investigated:
\begin{figure}[htbp]
\centering
\includegraphics[width=0.48\textwidth]{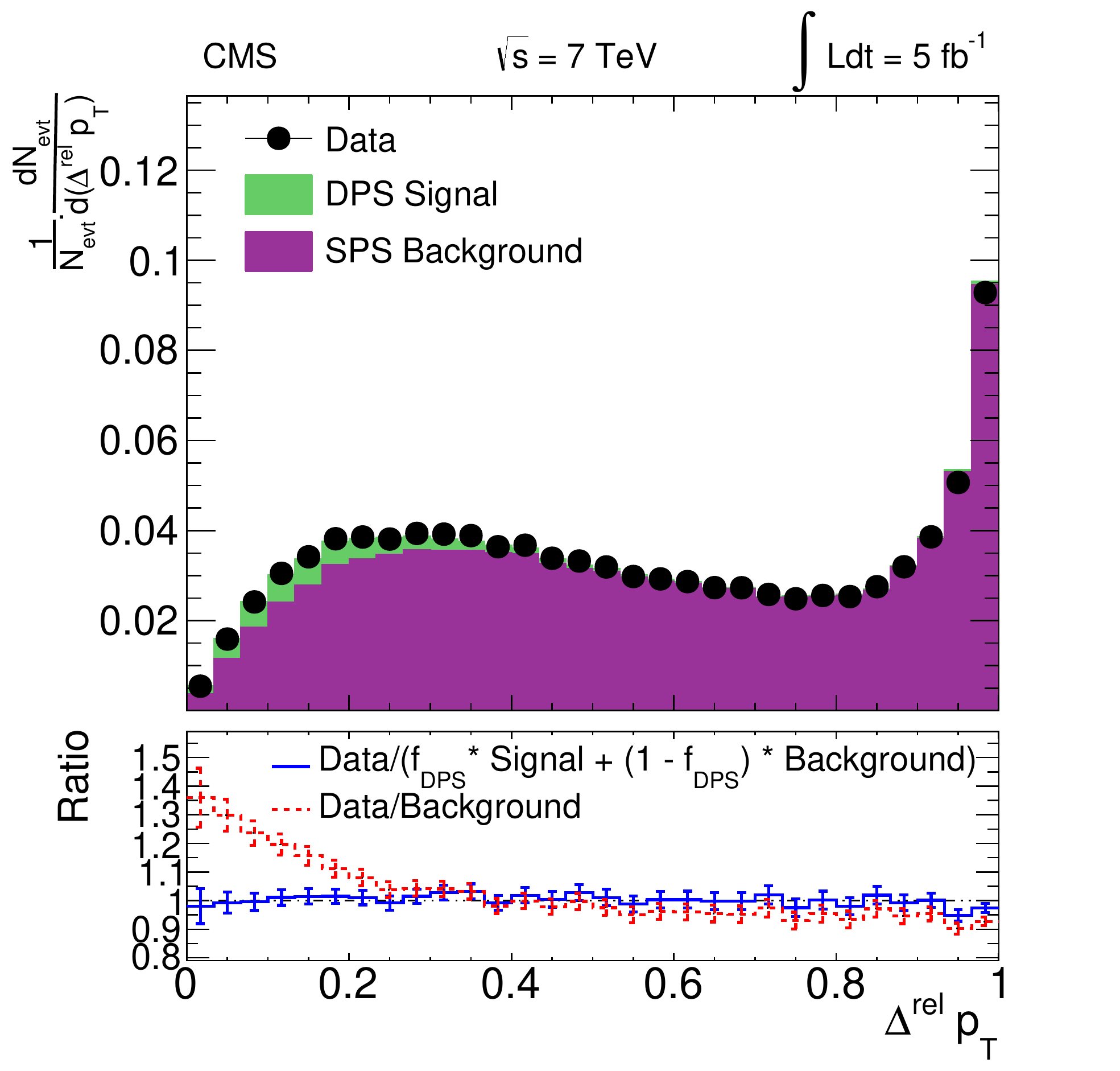}
\includegraphics[width=0.48\textwidth]{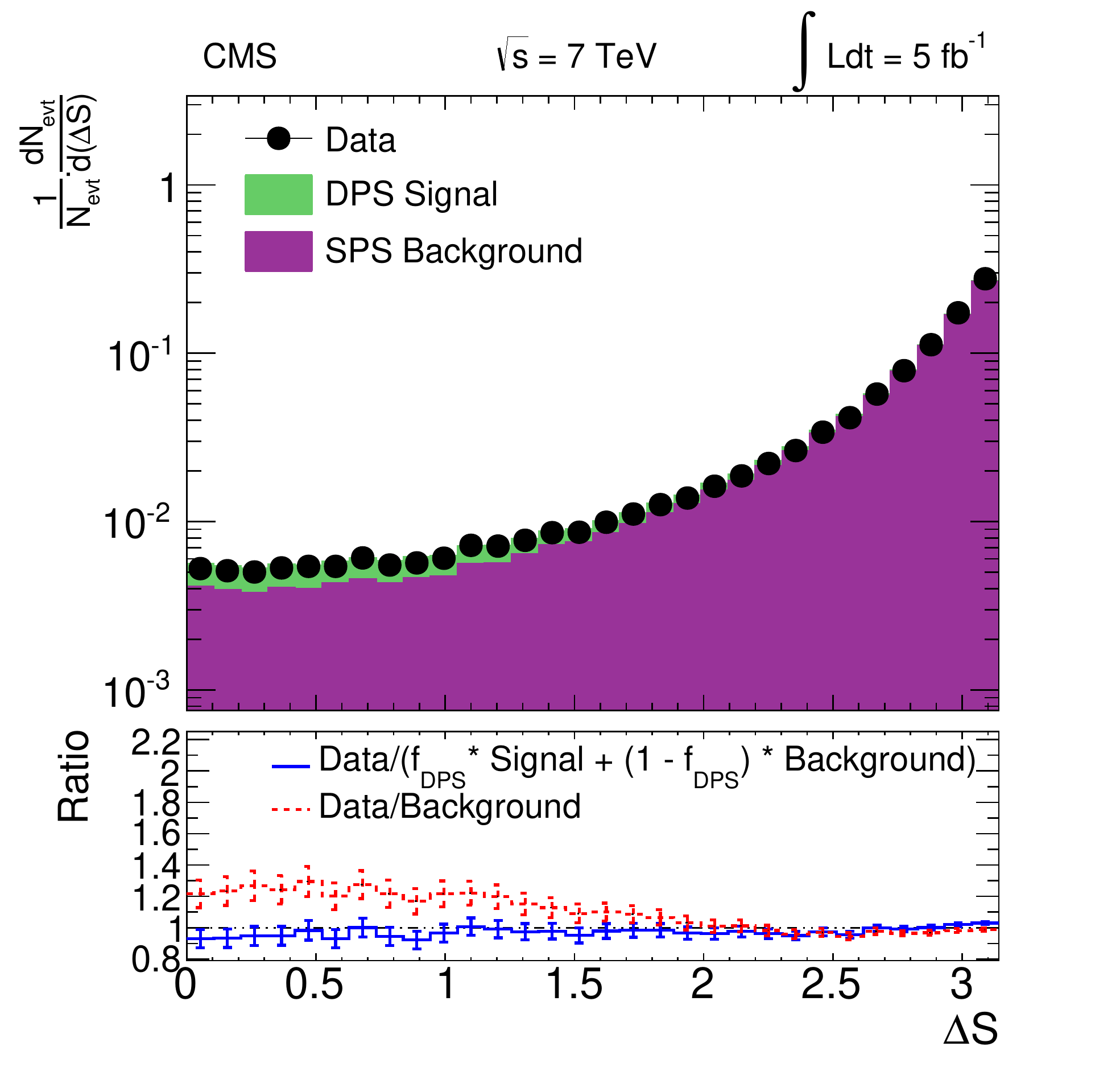}
\caption{Fit results for the DPS-sensitive observables $\Delta ^\text{rel}~{\pt}$ (left) and $\DS$ (right).
Corrected data distributions are fitted with signal and background templates (as discussed in section~\ref{sec:sigbkg}).
}
\label{fig:fit_data}
\end{figure}

\begin{itemize}
 \item
 the signal template is generated by randomly mixing W + 0-jet and dijet events from simulated events.
 The systematic uncertainty in the signal template is calculated by using different simulations for dijet events, i.e. \PYTHIA8, \POWHEG2, and \HERWIG{}++~\cite{hpp}.
 In this signal definition, the first and second interactions are assumed to be completely independent of each other.
 In order to study possible effects of colour reconnection and energy conservation between the first and the second interactions,
 an additional cross-check is performed by using the \PYTHIA8 event generator for producing W bosons from the first interaction and the dijet from the second interaction.
 From Fig.~\ref{fig:cor_norm} it has been concluded that \PYTHIA8 fails to describe W + 2-jet measurement due to missing contributions from 2$\rightarrow$3 and higher order processes.
 However, the DPS signal definition only includes W + 0-jet from the first interaction and exactly two jets from the second interaction, which are essentially 2$\rightarrow$1 and 2$\rightarrow$2 processes, respectively.
 Therefore, \PYTHIA8 gives a reliable simulation of the DPS signal events. 
 There is a significant contribution from the combinatorial background at particle level, i.e. at least one of the two jets is not from the second interaction.
 For systematic studies, this combinatorial background is removed by performing jet-parton matching.
 The variation in the signal template introduces a systematic uncertainty of 9\% in the extracted DPS fraction.
 MPI effects are included in both sets of events used for constructing the signal template; the effect of double counting is studied by switching off MPI in both sets of events, and found to be negligible.
 \item
 The background template is generated with \MADGRAPH5 events followed by hadronization and parton showering with the \PYTHIA8 event generator.
 In order to study the corresponding systematic uncertainty, the background template is obtained from various simulations and parton showering models, i.e. \MADGRAPH5 + \PYTHIA8, \MADGRAPH5 + \PYTHIA6 tune Z2*, and \POWHEG2 + \HERWIG6.
 The MPI partons cannot be tagged in \PYTHIA6 and \HERWIG6.
 The background template is therefore obtained by switching off MPI.
 The systematic uncertainty is evaluated by comparing the DPS fractions using different background templates with MPI off.
 In this procedure, the systematic uncertainty might be overestimated because of the construction of the background templates with simulations in which MPI are turned off.
 These variations in the background template introduce a systematic uncertainty of 17\% in the DPS fraction.

The systematic uncertainty related to the missing higher order diagrams in \MADGRAPH5 is estimated by varying the QCD renormalization and factorization scales simultaneously up and down by a factor of two.
This variation gives a systematic effect of 10\% on the extracted value of the DPS fraction.

The ME/PS matching scale is 20\GeV for the \MADGRAPH5 + \PYTHIA8 simulation.
This $\pt$-threshold controls the matching of partons produced from the matrix element and that from parton showering.
Systematic effects related to the matching scale are estimated by varying the $\pt$-threshold from 15\GeVc to 25\GeVc.
This variation gives a systematic uncertainty of 8\% in the DPS fraction.

The total systematic uncertainty on the background template is estimated to be 21\%.

 \item
 The effect of the uncertainty in PDFs used for the simulated sample is studied by using the PDF reweighting method and the prescription given in ref.~\cite{pdflhc}.
 The PDF reweighting only affects the hard scatterings and not the MPI and parton showers. The PDF uncertainties have little effect on the signal template but the variations in the background template result in an uncertainty of 5\% in the extracted DPS fraction.

 \item
 The systematic uncertainty due to the limited number of simulated events for the background template is obtained by varying the template within its statistical uncertainty.
 This gives a systematic uncertainty of 5\% in the extracted DPS fraction.

 \item
 The corrected measured distributions have various systematic uncertainties, as discussed in section~\ref{sec:correction}.
 The effect of these systematic biases is evaluated by varying the shape of the measured distributions within uncertainties. This variation affects the DPS fraction by 10\%.
 \end{itemize}

 Table~\ref{fDPSsys} summarizes the various systematic uncertainties in the extracted value of the DPS fraction.

\begin{table}[htb]
\centering
\topcaption{\label{fDPSsys} Systematic uncertainties in the DPS fraction determination.}
\begin{tabular}{l c}
\hline
 Source   & Uncertainty (\%) \\ \hline
 Signal template & 9 \\
 Background template & 21\\
 PDFs & 5\\
 Limited MC statistics & 5\\
 Uncertainty in corrected data & 10\\ \hline
 Total &  26\\
 \hline
\end{tabular}
\end{table}

\section{Determination of the effective cross section}
\label{sec:results}
As discussed in Section~\ref{sec:effxsec}, the effective cross section can be written as
\begin{equation}
\label{eq:effXsec}
 \sigma_\text{eff} = \frac{R}{f_\mathrm{DPS}}\cdot\sigma'_{2\rj}.
\end{equation}
To calculate the effective cross section, in addition to ${f_\mathrm{DPS}}$, the measurements of the dijet cross section and $R$ are also necessary.
They are discussed below.

\subsection{Measurement of \texorpdfstring{$R$}{R}}
The ratio, $R$, of the yield of events with a W boson in the final state and no jets to the yield of events with a W boson and exactly two jets with ${\pt} > 20$\GeVc and $\abs{\eta} < 2.0$ is $25.9\pm0.2\stat$ at detector level.
After subtracting the background contributions from the data, this ratio becomes $27.0\pm0.2\stat$.
The ratio $R$ is unfolded to particle level with a correction factor, ($R^\text{gen}/R^\text{reco}$), of 1.03 is calculated with \MADGRAPH5 + \PYTHIA6.
The corrected value of $R$ is 27.8 with a statistical uncertainty of 0.7\%.
The measurement of $R$ has a total systematic uncertainty of 12\% due to various sources, as listed in Table~\ref{Rsys}.
The measured value of $R$ is:

\begin{equation}
R = 27.8 \pm 0.2\stat\pm 3.3\syst.
\end{equation}

\begin{table}[htb]
 \centering
\caption{\label{Rsys} Systematic uncertainties in the measurement of $R$.}
\begin{tabular}{l c}
\hline
 Source   & Uncertainty (\%) \\ \hline
 Model dependence & 9 \\
 JES & 7\\
 JER & 2\\
 Background & 2\\
 Pileup & 1\\ \hline
 Total & 12\\
 \hline
\end{tabular}
\end{table}

\subsection{Measurement of \texorpdfstring{$\sigma'_{2\rj}$}{TEST}}

The cross section for dijet production with ${\pt} > 20$\GeVc and $\abs{\eta} < 2.0$ ($ \sigma'_{2\rj}$) is measured with pp collision data at a centre-of-mass energy of 7\TeV collected during 2010.
A combination of minimum bias and single-jet triggers is used, as for the inclusive jet cross section measurement~\cite{jetXsec}.
For each trigger, the offline jet ${\pt}$ threshold is chosen such that the trigger is 100\% efficient.

At detector level, the cross section for events with exactly two jets with ${\pt} > 20$\GeVc and $\abs{\eta} < 2.0$ is measured to be 0.046 mb.
This is corrected to particle level with a correction factor of 0.89 calculated from the \PYTHIA6 simulation.
The 8\% uncertainty in the correction factor is due to the model dependence, estimated by comparing the corrections obtained from the \PYTHIA6 and \HERWIG{}++ samples.
There are further systematic uncertainties of 13\% and 2\% due to the JES and JER uncertainties, respectively.
Table~\ref{two-jetsys} summarizes the various sources of systematic uncertainties.
The production cross section $\sigma'_{2\rj}$ at particle level is:

\begin{equation}
\sigma'_{2\rj} = 0.0409 \pm 0.0004\stat \pm 0.0061\syst\unit{mb}.
\end{equation}

\begin{table}[htb]
\centering
\topcaption{\label{two-jetsys} Uncertainties in the measured value of the dijet cross section.}
\begin{tabular}{l c}
\hline
 Source   & Uncertainty (\%) \\ \hline
 Model dependence & 8 \\
 JES & 13\\
 JER & 2\\ \hline
 Total & 15\\
 \hline
\end{tabular}
\end{table}

With the values of ${f_\mathrm{DPS}}$, R, and ${\sigma'_{2\rj}}$ in eq.(\ref{eq:effXsec}), the effective cross section is determined to be:

\begin{equation}
\sigma_\text{eff} = 20.7 \pm 0.8\stat \pm  6.6\syst\unit{mb}.
\end{equation}

\begin{figure}[htbp]
\centering
\includegraphics[width=0.65\textwidth]{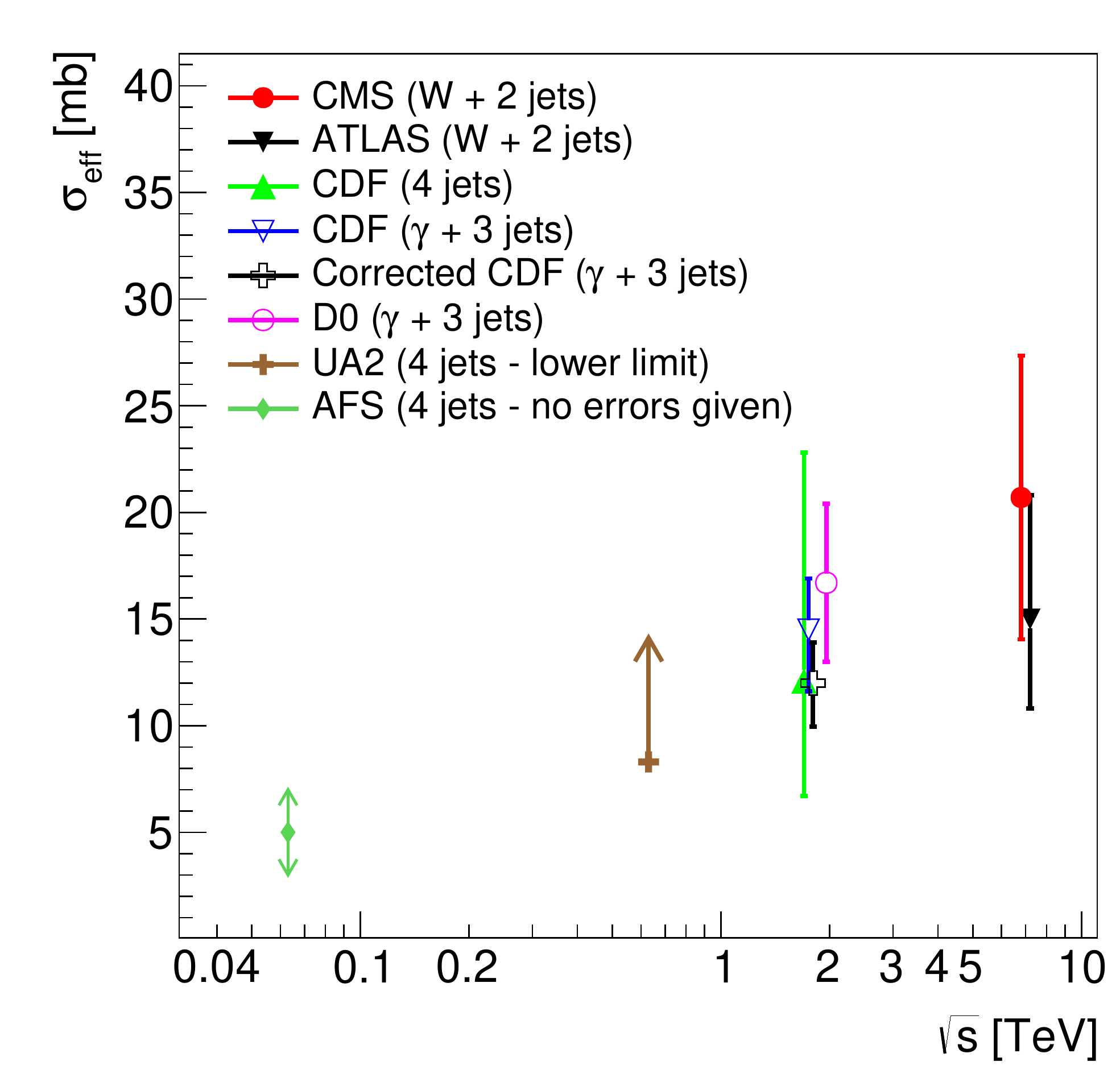}
\caption{Centre-of-mass energy dependence of $\sigma_\text{eff}$ measured by different experiments using different processes~\cite{UA2,AFS,CDF1,CDF2,D0,atlas}.
These measurements used different approaches for extraction of the DPS fraction and $\sigma_\text{eff}$.
The ``Corrected CDF'' data point indicates the $\sigma_\text{eff}$ value corrected for the exclusive event selection~\cite{Seymour1}.
}
\label{fig:com}
\end{figure}

The results of the measurements of R, the DPS fraction, the dijet cross section, and the effective cross section are summarized in Table~\ref{summary}.
Figure~\ref{fig:com} shows a comparison of the effective cross sections obtained using different processes at various centre-of-mass energies.
In some theoretical models~\cite{Treleani}, $\sigma_\text{eff}$ is expressed as a simple geometrical integral that is independent of the collision energies.
Conversely, \PYTHIA predicts an increase of $\sigma_\text{eff}$  with centre-of-mass energy because of the parameter ${{\pt}_0} (\sqrt{s})$, which depends on the collision energy.
From the experimental results, a firm conclusion on the energy dependence of $\sigma_\text{eff}$ cannot be drawn because of the large systematic uncertainties.
The CMS measurement is consistent with previous measurements performed at the Tevatron and by the ATLAS Collaboration at the LHC.
The CMS measurement is also consistent with predictions from \PYTHIA of 20--30\unit{mb}, depending on the tune.
The measured effective cross section is also consistent with that obtained by fitting the underlying event data~\cite{effXsecUE} with the \HERWIG{}++ simulation.

\begin{table}
\centering
\topcaption{\label{summary} Measured value of  ${f_\mathrm{DPS}}$, R, $\sigma'_{2\rj}$, and the effective cross section.}
\begin{tabular}{l c}
\hline
 ${f_\mathrm{DPS}}$ &  0.055 $\pm$ 0.002\stat $\pm$ 0.014\syst\\
 $R$ & 27.8 $\pm$ 0.2\stat $\pm$ 3.3\syst\\
 $\sigma'_{2\rj}$ & 0.0409 $\pm$ 0.0004\stat $\pm$ 0.0061\syst mb\\
 Effective cross section & 20.7 $\pm$ 0.8\stat $\pm$  6.6\syst mb\\
 \hline
\end{tabular}
\end{table}

\section{Summary}
A study of double parton scattering in W + 2-jet events in pp collisions is presented.
The data correspond to an integrated luminosity of 5 \fbinv collected in pp collisions at a centre-of-mass energy of 7\TeV.
Events with a W boson, reconstructed from the muon and \ETslash information,  are required to have exactly two jets with ${\pt} > 20$\GeVc and  $\abs{\eta} < 2.0$.
The DPS-sensitive observables $\Delta ^\text{rel}~{\pt}$ and $\Delta S$ are corrected for detector effects and selection efficiencies.
Simulations of W + jets events with \MADGRAPH5 + \PYTHIA8 (or \PYTHIA6) and NLO predictions of \POWHEG2 + \PYTHIA6 (or \HERWIG6) provide a good description of the observables and describe the data only if multiple parton
interactions are included.

The fraction of DPS in W + 2-jet events is extracted with a DPS + SPS  template fit to the distribution of the $\Delta ^\text{rel}~{\pt}$ and $\DS$ observables.
The obtained value of the DPS fraction is

\begin{equation}
{f_\mathrm{DPS}} = 0.055 \pm 0.002\stat \pm 0.014\syst,
\end{equation}

and the effective cross section, characterizing the effective transverse area of hard partonic interactions in collisions between protons, is calculated to be

\begin{equation}
\sigma_\text{eff} = 20.7 \pm 0.8\stat \pm  6.6\syst\unit{mb}.
\end{equation}

The measured value of the effective cross section is consistent with the Tevatron and ATLAS results.

\section*{Acknowledgements}
We would like to thank M.~Diehl, P.~Nason, M.H.~Seymour, T.~Sj{\"o}strand, P.~Skands and D.~Treleani for key suggestions related to the theoretical interpretation of the measurement.
\hyphenation{Bundes-ministerium Forschungs-gemeinschaft Forschungs-zentren} We congratulate our colleagues in the CERN accelerator departments for the excellent performance of the LHC and thank the technical and administrative staffs at CERN and at other CMS institutes for their contributions to the success of the CMS effort. In addition, we gratefully acknowledge the computing centres and personnel of the Worldwide LHC Computing Grid for delivering so effectively the computing infrastructure essential to our analyses. Finally, we acknowledge the enduring support for the construction and operation of the LHC and the CMS detector provided by the following funding agencies: the Austrian Federal Ministry of Science and Research and the Austrian Science Fund; the Belgian Fonds de la Recherche Scientifique, and Fonds voor Wetenschappelijk Onderzoek; the Brazilian Funding Agencies (CNPq, CAPES, FAPERJ, and FAPESP); the Bulgarian Ministry of Education and Science; CERN; the Chinese Academy of Sciences, Ministry of Science and Technology, and National Natural Science Foundation of China; the Colombian Funding Agency (COLCIENCIAS); the Croatian Ministry of Science, Education and Sport, and the Croatian Science Foundation; the Research Promotion Foundation, Cyprus; the Ministry of Education and Research, Recurrent financing contract SF0690030s09 and European Regional Development Fund, Estonia; the Academy of Finland, Finnish Ministry of Education and Culture, and Helsinki Institute of Physics; the Institut National de Physique Nucl\'eaire et de Physique des Particules~/~CNRS, and Commissariat \`a l'\'Energie Atomique et aux \'Energies Alternatives~/~CEA, France; the Bundesministerium f\"ur Bildung und Forschung, Deutsche Forschungsgemeinschaft, and Helmholtz-Gemeinschaft Deutscher Forschungszentren, Germany; the General Secretariat for Research and Technology, Greece; the National Scientific Research Foundation, and National Innovation Office, Hungary; the Department of Atomic Energy and the Department of Science and Technology, India; the Institute for Studies in Theoretical Physics and Mathematics, Iran; the Science Foundation, Ireland; the Istituto Nazionale di Fisica Nucleare, Italy; the Korean Ministry of Education, Science and Technology and the World Class University program of NRF, Republic of Korea; the Lithuanian Academy of Sciences; the Mexican Funding Agencies (CINVESTAV, CONACYT, SEP, and UASLP-FAI); the Ministry of Business, Innovation and Employment, New Zealand; the Pakistan Atomic Energy Commission; the Ministry of Science and Higher Education and the National Science Centre, Poland; the Funda\c{c}\~ao para a Ci\^encia e a Tecnologia, Portugal; JINR, Dubna; the Ministry of Education and Science of the Russian Federation, the Federal Agency of Atomic Energy of the Russian Federation, Russian Academy of Sciences, and the Russian Foundation for Basic Research; the Ministry of Education, Science and Technological Development of Serbia; the Secretar\'{\i}a de Estado de Investigaci\'on, Desarrollo e Innovaci\'on and Programa Consolider-Ingenio 2010, Spain; the Swiss Funding Agencies (ETH Board, ETH Zurich, PSI, SNF, UniZH, Canton Zurich, and SER); the National Science Council, Taipei; the Thailand Center of Excellence in Physics, the Institute for the Promotion of Teaching Science and Technology of Thailand, Special Task Force for Activating Research and the National Science and Technology Development Agency of Thailand; the Scientific and Technical Research Council of Turkey, and Turkish Atomic Energy Authority; the Science and Technology Facilities Council, UK; the US Department of Energy, and the US National Science Foundation.
Individuals have received support from the Marie-Curie programme and the European Research Council and EPLANET (European Union); the Leventis Foundation; the A. P. Sloan Foundation; the Alexander von Humboldt Foundation; the Belgian Federal Science Policy Office; the Fonds pour la Formation \`a la Recherche dans l'Industrie et dans l'Agriculture (FRIA-Belgium); the Agentschap voor Innovatie door Wetenschap en Technologie (IWT-Belgium); the Ministry of Education, Youth and Sports (MEYS) of Czech Republic; the Council of Science and Industrial Research, India; the Compagnia di San Paolo (Torino); the HOMING PLUS programme of Foundation for Polish Science, cofinanced by EU, Regional Development Fund; and the Thalis and Aristeia programmes cofinanced by EU-ESF and the Greek NSRF.

\newpage

\bibliography{auto_generated}

\cleardoublepage \appendix\section{The CMS Collaboration \label{app:collab}}\begin{sloppypar}\hyphenpenalty=5000\widowpenalty=500\clubpenalty=5000\textbf{Yerevan Physics Institute,  Yerevan,  Armenia}\\*[0pt]
S.~Chatrchyan, V.~Khachatryan, A.M.~Sirunyan, A.~Tumasyan
\vskip\cmsinstskip
\textbf{Institut f\"{u}r Hochenergiephysik der OeAW,  Wien,  Austria}\\*[0pt]
W.~Adam, T.~Bergauer, M.~Dragicevic, J.~Er\"{o}, C.~Fabjan\cmsAuthorMark{1}, M.~Friedl, R.~Fr\"{u}hwirth\cmsAuthorMark{1}, V.M.~Ghete, C.~Hartl, N.~H\"{o}rmann, J.~Hrubec, M.~Jeitler\cmsAuthorMark{1}, W.~Kiesenhofer, V.~Kn\"{u}nz, M.~Krammer\cmsAuthorMark{1}, I.~Kr\"{a}tschmer, D.~Liko, I.~Mikulec, D.~Rabady\cmsAuthorMark{2}, B.~Rahbaran, H.~Rohringer, R.~Sch\"{o}fbeck, J.~Strauss, A.~Taurok, W.~Treberer-Treberspurg, W.~Waltenberger, C.-E.~Wulz\cmsAuthorMark{1}
% ...
\vskip\cmsinstskip
% ...
1:~~Also at Vienna University of Technology, Vienna, Austria\\
2:~~Also at CERN, European Organization for Nuclear Research, Geneva, Switzerland\\
% ...

\end{sloppypar}
\end{document}